\documentclass[%
 reprint,
superscriptaddress,
 amsmath,amssymb,
 aps,
 prx,
]{revtex4-2}

\usepackage{graphicx}
\usepackage{dcolumn}
\usepackage{bm}
\usepackage{svg}
\svgpath{{../Figures/Fridge/}}
\usepackage{microtype} 
\usepackage{paralist}
\usepackage[normalem]{ulem}
\usepackage[unicode=true, colorlinks=true, citecolor={blue!80!black}, urlcolor={blue!50!black}, linkcolor = {blue!80!black}]{hyperref} 
\usepackage{array}
\usepackage{siunitx}
\usepackage{wrapfig}
\usepackage{enumitem}
\usepackage{amsmath,amssymb}
\usepackage{array}
\usepackage{soul}
\usepackage{centernot}
\usepackage{bibunits}
\usepackage[titletoc, title]{appendix}
\usepackage{titletoc}
\usepackage{lineno}
\usepackage{lipsum}  
\usepackage{mathtools}
\usepackage{tabularx}

\newcommand*{\balancecolsandclearpage}{%
  \cleardoublepage
  \twocolumngrid
}

\DeclareSIUnit\angstrom{\text {Å}}

\begin{document}

\preprint{APS/123-cQED}

\title{
Low-loss Nb on Si superconducting resonators from a dual-use spintronics deposition chamber and with acid-free post-processing
}

\author{Maciej W. Olszewski}
\email{mwo34@cornell.edu}
\affiliation{Department of Physics, Cornell University, Ithaca, NY 14853, USA}

\author{Jadrien T. Paustian}
\affiliation{Department of Physics, Syracuse University, NY 13244-1130, USA}

\author{Tathagata Banerjee}
\affiliation{School of Applied and Engineering Physics, Cornell University, Ithaca, NY 14853, USA}

\author{Haoran Lu}
\affiliation{School of Applied and Engineering Physics, Cornell University, Ithaca, NY 14853, USA}

\author{Jorge L. Ramirez}
\affiliation{Department of Physics, University of Colorado Boulder, Boulder, CO 80309, USA}
\affiliation{National Institute for Standards and Technology Boulder, Boulder, CO 80305, USA}

\author{Nhi Nguyen}
\affiliation{Department of Physics, University of Colorado Boulder, Boulder, CO 80309, USA}
\affiliation{National Institute for Standards and Technology Boulder, Boulder, CO 80305, USA}

\author{Kiichi Okubo}
\affiliation{Department of Physics, Syracuse University, NY 13244-1130, USA}

\author{Rohit Pant}
\affiliation{Laboratory for Physical Sciences, 8050 Greenmead Dr, College Park, MD 20740, USA}

\author{Aleksandra B. Biedron}
\affiliation{NY Creates, Albany, NY 12203, USA}

\author{Daniel C. Ralph}
\affiliation{Department of Physics, Cornell University, Ithaca, NY 14853, USA}
\affiliation{Kavli Institute at Cornell, Ithaca, New York 14853, USA}

\author{Christopher J. K. Richardson}
\affiliation{Laboratory for Physical Sciences, 8050 Greenmead Dr, College Park, MD 20740, USA}
\affiliation{Department of Materials Science and Engineering, University of Maryland, College Park, MD 20742, USA}

\author{Gregory D. Fuchs}
\affiliation{School of Applied and Engineering Physics, Cornell University, Ithaca, NY 14853, USA}

\author{Corey Rae H. McRae}
\affiliation{Electrical, Computer, and Energy Engineering Department, University of Colorado Boulder, Boulder, CO 80309, USA}
\affiliation{Department of Physics, University of Colorado Boulder, Boulder, CO 80309, USA}
\affiliation{National Institute for Standards and Technology Boulder, Boulder, CO 80305, USA}

\author{Ivan V. Pechenezhskiy}
\affiliation{Department of Physics, Syracuse University, NY 13244-1130, USA}

\author{B. L. T. Plourde}
\altaffiliation{Present affiliation: University of Wisconsin-Madison}  
\affiliation{Department of Physics, Syracuse University, NY 13244-1130, USA}

\author{Valla Fatemi}
\email{vf82@cornell.edu}
\affiliation{School of Applied and Engineering Physics, Cornell University, Ithaca, NY 14853, USA}

\date{\today}


\begin{abstract} 
Magnetic impurities are known to degrade superconductivity. 
For this reason, physical vapor deposition chambers that have previously been used for magnetic materials have generally been avoided for making high-quality superconducting resonator devices. 
In this article, we show by example that such chambers can be used for this purpose; with Nb films sputtered in a chamber that continues to be used for magnetic materials, we demonstrate compact (\SI{3}{\micro\meter} gap) coplanar waveguide resonators with low-power internal quality factors near one million.
We achieve this using a resist strip bath with no post-fabrication acid treatment, which results in performance comparable to previous strip baths with acid treatments.
We also find evidence that this improved resist strip bath provides a better surface chemical template for post-fabrication hydrogen fluoride processing.
These results are consistent across three Si substrate preparation methods, including a \SI{700}{\celsius} anneal.
These results will inform nanofabrication for other superconducting materials and the integration of magnetic materials for hybrid systems.
\end{abstract}

\maketitle

\section{Introduction}
Superconducting microwave circuits are one of the leading candidate platforms for enabling quantum computing technology~\cite{kjaergaard_superconducting_2020,sivak_real-time_2023,acharya_quantum_2024}, and Nb on Si is a commonly implemented materials set~\cite{mcrae_materials_2020}.
A promising avenue for improving device performance and technological prospects is to eliminate defects from the materials and interfaces within the superconducting circuits~\cite{mcrae_materials_2020,de_leon_materials_2021}. 

Surface treatments~\cite{woods_determining_2019,oconnell_microwave_2008,earnest_substrate_2018,ganjam_surpassing_2024,crowley_disentangling_2023,read_precision_2022,kopas_characterization_2020,megrant_planar_2012,murthy_tof-sims_2022,lee_discovery_2023} and deposition conditions~\cite{kopas_characterization_2020,drimmer_effect_2024,lee_discovery_2023,oh_exploring_2024} have been demonstrated to have significant effects on superconducting device performance. 
Of particular concern are magnetic impurities, which can rapidly suppress critical temperature~\cite{anderson_theory_1959}, introduce flux noise~\cite{silva_subgap_2005,sendelbach_magnetism_2008,paladino_mathbsf1mathbsfitf_2014,muller_towards_2019}, and induce sub-gap density of states in the superconductor which can result in internal losses~\cite{woolf_effect_1965,balatsky_lifshitz_1997,fominov_subgap_2016}.
These impurities can be introduced to superconducting quantum devices by the deposition chamber environment if a magnetic material target is present, or if magnetic materials have been previously deposited. However, the environmental threshold at which device performance is impacted has not been established.

Here, we report that a high vacuum magnetron sputter chamber in active use for depositing magnetic materials can be concurrently used for synthesis of superconducting thin films that produce coplanar microwave resonators with high internal quality factors $Q_i$ at single-photon powers. 
We focus on Nb films sputtered on high-resistivity Si (100) substrates, and perform optimization of pre-deposition substrate preparation methods and post-fabrication surface treatments. 
Materials analysis indicates no detectable magnetic impurities in the film or substrate.
These results indicate that materials and synthesis exploration can be accomplished more broadly and inexpensively across the community, without the requirement of deposition systems dedicated uniquely to superconducting circuits.
Further, we establish a method for fabricating high-quality devices with acid-free post-processing, and additionally find indications that surface quality can impact the level of improvement provided by acid-free postprocessing.
This may be useful for qubit fabrication as some materials, including conventional Josephson junctions, are attacked by the acids often used in post-processing of refractory metals like Nb and Ta.

\section{Chamber hygiene and fabrication }
\subsection{Chamber Hygiene} \label{sec:hygiene}
This case study focuses on an AJA International, Inc. magnetron sputtering chamber which has primarily been used for magnetic depositions for more than 20  years~\cite{fuchs_spin-transfer_2004}, including iron, cobalt, and nickel.
The chamber is equipped with seven two-inch gun assemblies and can reach base pressures of $3\times10^{-8}$ Torr without gettering. 
This study includes both separate and co-loading of superconducting and magnetic deposition targets through two slightly different cleaning and maintenance procedures, as described in detail in Appendix~\ref{app:chamber}.
We refer to them here as deposition procedure 1 (DP1), which was used when the Nb sputter target was co-loaded with only other superconducting material targets, and deposition procedure 2 (DP2), which was used when the Nb sputter target was co-loaded with magnetic targets, including permalloy (Ni$_{80}$Fe$_{20}$), cobalt (Co), and cobalt-iron-boron (Co$_{20}$Fe$_{60}$B$_{20}$).
For the case of DP2, the magnetic targets were frequently used for other depositions.
DP1 is featured in both datasets below (Fig.~\ref{fig:boxplots} and Fig.~\ref{fig:boxplots_new}) while DP2 is only in the second group (Fig.~\ref{fig:boxplots_new}). 
The two procedures did not significantly impact our device performance, as will be discussed below in Sec.~\ref{sec:discussion}. 

\subsection{Wafer preparation methods}
To demonstrate the range of device performances emerging from a dual-use spintronics chamber, we investigate three methods of preparing the substrate before deposition: \textbf{BOE}, \textbf{Anneal}, and \textbf{Thermal}.
All of them involve a BOE dip immediately prior to loading into the chamber to remove the silicon surface oxide. 
The Anneal and Thermal preparations additionally involve a flash anneal at \SI{700}{\celsius} in the deposition chamber. 
The Thermal preparation separately involves processing that removes the top $\sim\SI{100}{\nano\meter}$ of the wafer.
The processes are described in detail in Appendix~\ref{app:substrate}.
All depositions are performed on \SI{525}{\micro\meter} thick 100-mm float-zone silicon (100) wafers with resistivity $\geq \SI{10}{\kilo\ohm\centi\meter}$ (WaferPro).

\begin{figure*}
\centering
\includegraphics[width = 1.95\columnwidth]{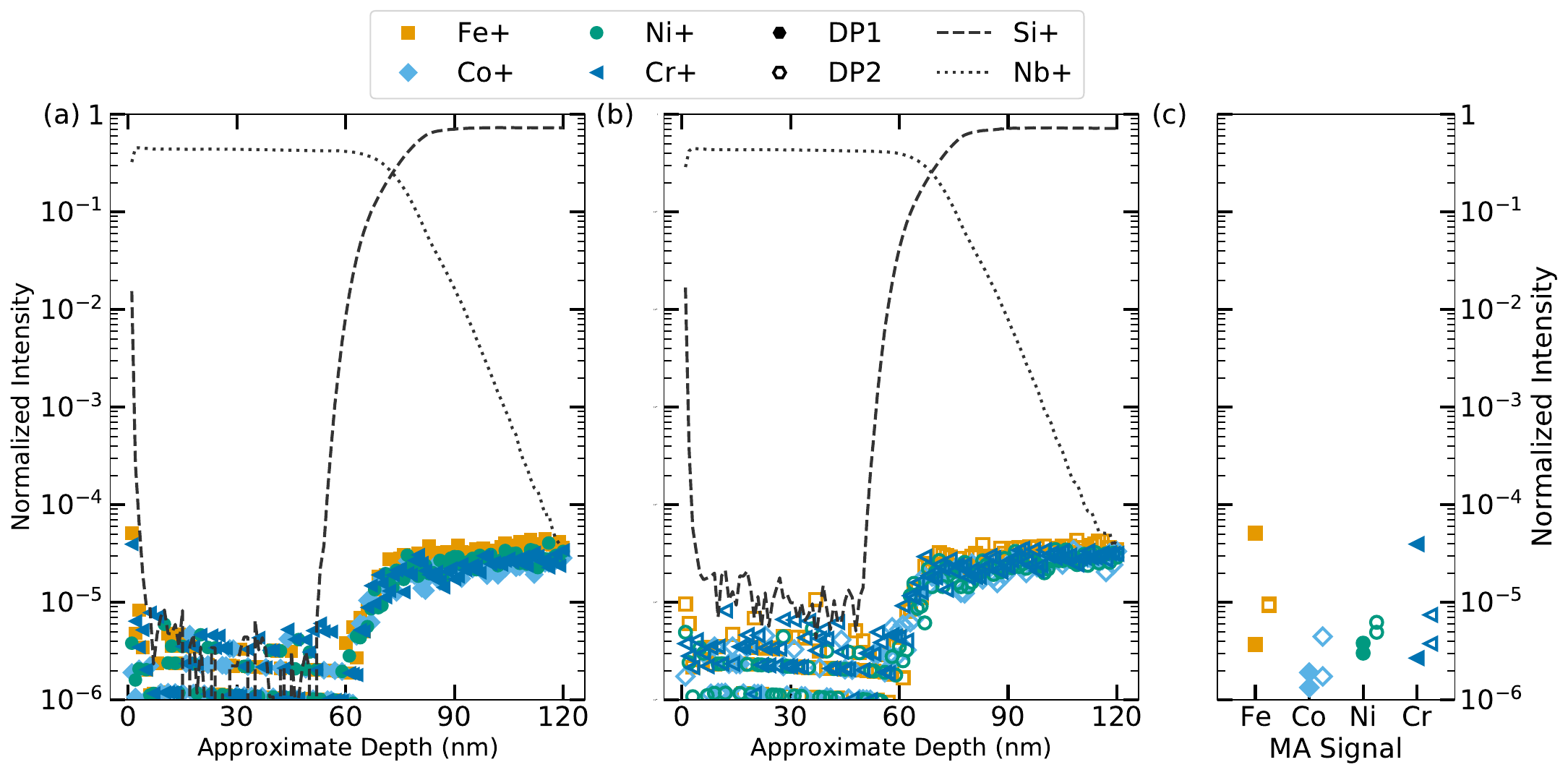}
\caption{
Secondary-ion mass spectrometry (SIMS) profiles of magnetic impurities (Fe, Co, Ni, and Cr) as a function of depth in a \SI{60}{\nano\meter}  Nb film.
The raw intensity of impurity ions was normalized to the total secondary ion counts.
(a) A Nb film deposited with only superconductors present in the chamber (DP1).
(b) A Nb film deposited with co-loaded magnetic targets having been used in the chamber (DP2).
In both cases, no magnetic contamination is detected in the bulk of the film or at the substrate-metal interface. 
Differences in signal are only observed at the metal-air (MA) interface.
(c) Additional measurements focused on the metal-air interface of similar Nb samples for both the control and DP2 films.
Solid filled shapes represent control samples and unfilled shapes represent DP2 samples.
}
\label{fig:sims_magnet} 
\end{figure*}

\subsection{Key device fabrication aspects}
We deposit \SI{60}{\nano\meter} Nb films at \SI{2.4}{\angstrom\per\second} following each of the three substrate preparation methods. 
Here we highlight the key steps of the fabrication process, with details given in Appendix~\ref{app:fabrication}.

After photolithography and dry etching of the Nb with a chlorine-based etch chemistry, the photo-resist is removed by a heated soak in a resist strip bath. 
Two baths were used: MICROPOSIT 1165 resist remover, hereinafter called \textbf{1165}, or Integrated Micro Materials AZ 300T stripper, referred to as \textbf{AZ}.
The 1165 bath was followed by a series of sonications in 1165, acetone, isopropanol, and finishing with a oxygen plasma descum.
The AZ bath was followed by sonications in isopropanol and deionized (DI) water, and rinses in DI water, and isopropanol, finishing with an oxygen plasma descum.

Certain devices were dipped in a 2\% HF solution for either 18 or 60 seconds, a few hours up to a couple of days before loading for measurements as described in Table~\ref{tab:fridges}.
We note that this post-fabrication HF dose is intended to remove surface defects from the devices (e.g., any remaining magnetic contaminants and re-deposited Nb), and is not strong enough to fully remove the niobium surface oxide~\cite{torres-castanedo_formation_2024}.

\subsection{Materials characterization}

We examined our films with secondary-ion mass spectrometry (SIMS), atomic force microscopy (AFM), X-ray photoelectron spectroscopy (XPS), and cryogenic low-frequency electronic transport measurements. 

The superconducting transition temperature $T_c$ for the as-grown \SI{60}{\nano\meter} Nb films is above \SI{9.1}{\kelvin}, and the residual-resistance ratio (RRR) is above 4.5, comparable with sputtered Nb films reported in literature~\cite{anferov_improved_2023,drimmer_effect_2024,zheng_nitrogen_2022,van_damme_argon-milling-induced_2023}. 
The temperature dependencies of microwave resonator loss, as shown in Figure~\ref{fig:temperature}, also indicate values of $T_{c}$ near \SI{9}{\kelvin}.
The details of the AFM (App.~\ref{app:afm}) and the XPS (App.~\ref{app:xps}) characterizations are in the appendices.

We comment in particular on the SIMS measurements shown in Figure~\ref{fig:sims_magnet}, where we identify levels of the most commonly deposited magnetic elements in the tool, namely Fe, Co, Ni, and Cr, which are also low-vapor-pressure materials.
Measurements reveal no detectable impurities in the bulk of the Nb film and at the substrate-metal interface.
A small amount of Fe and Cr is detected at the metal-air interface, likely attached to the surface oxide rather than to the metallic Nb.
The presence of Fe is not consistent across all films measured and may result from handling after deposition or nanoscopic magnetic dust~\cite{oldfield_magnetic_1985}.
In particular, we note that SIMS did not detect magnetic impurities in any of the films deposited with the magnetic targets present (DP2) [Figure~\ref{fig:sims_magnet}(b)].
These surface magnetic impurities were also detected on films deposited in other, superconductor-only deposition chambers.

\begin{figure*}
\centering
\includegraphics[width = 1.95\columnwidth]{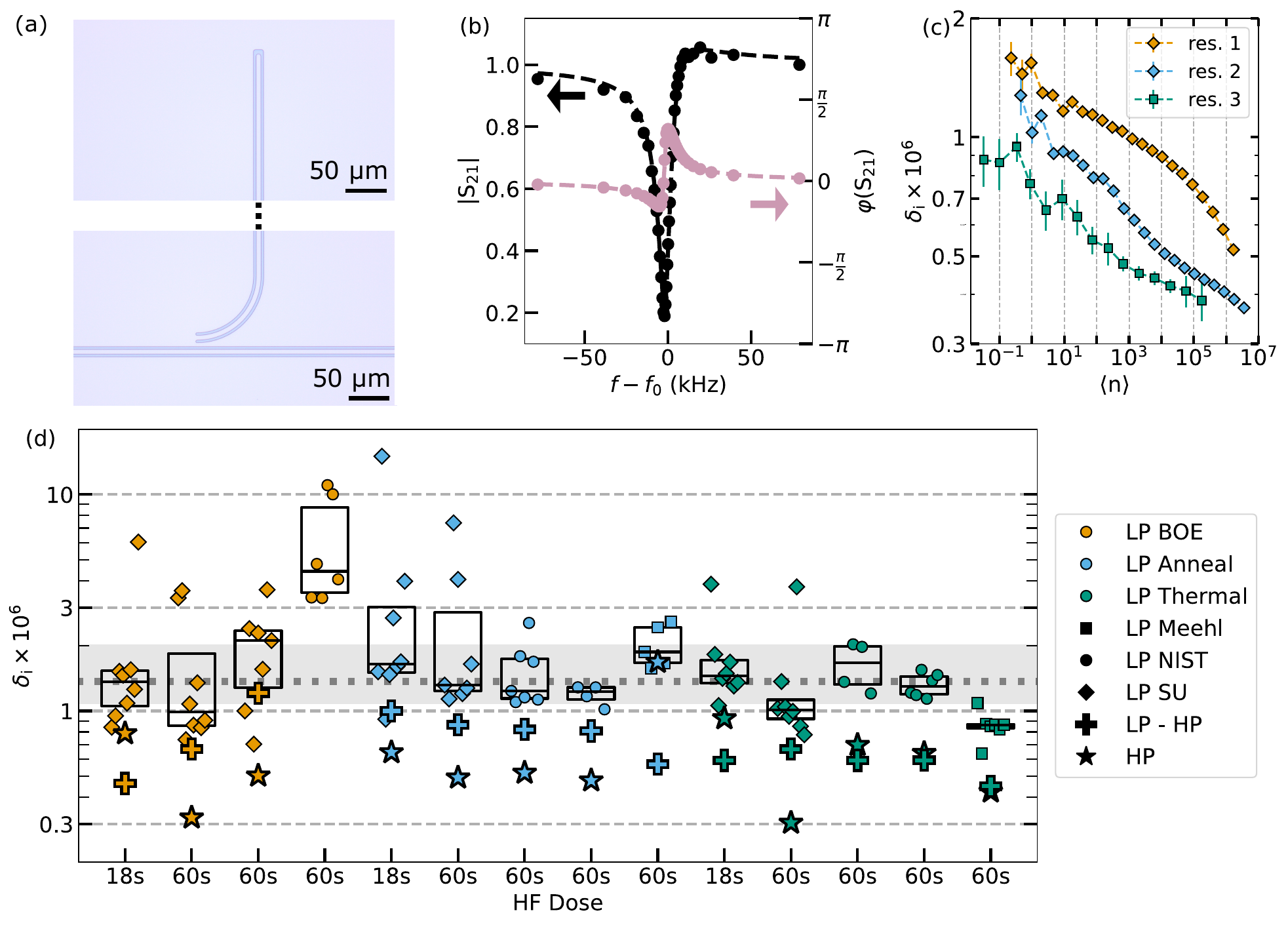}
\caption{
Summary of resonator measurements from DP1 performed across three measurement setups, referred to as Meehl, NIST, and SU for different substrate preparation methods with the 1165 sample treatment.
(a) Optical microscope device image of CPW resonator.
(b) A sample Thermal resonator measurement at $\langle n\rangle \approx 10^5$ photons on the Meehl cryostat at \SI{17}{\milli\kelvin}. 
Black and purple points are the measured magnitude and phase of $S_{21}$, respectively, plotted on the left and right axis, indicated by the arrows. 
The dashed lines are the fit for the data, from equation~\ref{eq:S21-fit}. 
(c) Three examples of loss $\delta_{i}$ vs. average photon number $\langle n\rangle$  for different samples.
Colors and shapes represent the sample type and cryostat used for measurements, respectively, as shown in the panel (d) legend.
(c) Box plots of low power losses $\delta_{LP}$ for the devices measured with the different substrate preparation methods. 
Colors and shapes represent the substrate preparation and the fridge, respectively.
HF dose refers to the duration of the post-fabrication HF treatment.
Star symbols plot the median HP losses for each device.
Plus symbols plot the median difference between LP and HP losses for each device.
HP measurements were not taken for the fourth device from the left.
Some combinations of preparation and measurement protocols were repeated for more than one sample die.
The shaded band is the window of 25th to 75th percentile performance of all of the resonators plotted, with the dashed line being the median.
}
\label{fig:boxplots}
\end{figure*}

\section{Microwave Resonator Results}

\subsection{Si preparation series}

We use the open-source resonator design developed by the Boulder Cryogenic Quantum Testbed to characterize the microwave properties of resonators fabricated from these films~\cite{kopas_simple_2022}.
This design has a \SI{3}{\micro\meter} gap between the center conductor and the ground plane and has a hangar resonator configuration, with eight $\lambda/4$ resonators at frequencies between 4 and \SI{8}{\giga\hertz}.
For each resonance, we locally vary the probe frequency and fit the complex-valued transfer function to the expected resonance response, as in Figure~\ref{fig:boxplots}(b). 
Accomplishing this as a function of drive power, and using the measured attenuation of the input coaxial lines, we can estimate the average photon number. 
We fit the value of the internal quality factor $Q_{i}$ (details in Appendix~\ref{app:res_fits}) and calculate the loss ($\delta_{i}$):
\begin{equation}
    \delta_{i}\approx\tan\delta_{i}=\frac{1}{Q_{i}}.
\end{equation}
For comparison across dies and preparation methods, we extract the experimentally-observed low power (LP) losses (average resonator loss for $\langle n \rangle < 1$) and high power (HP) losses (average resonator loss for $\langle n \rangle > 10^5$).

In Figure~\ref{fig:boxplots}(c), we show the internal loss of a resonator as a function of the average photon number in the resonator for devices corresponding to our three substrate preparation methods.  
The three sample resonators plotted are representative of the qualitatively different trends observed.
In particular, the data from resonator 2 and resonator 3 show signs of a `double-S' shaped curve, possibly indicative of two TLS (two level system) baths existing at different critical photon numbers~\cite{verjauw_investigation_2021}. This suggests that a usual TLS-based model is unsuitable for our data.
Note that the data from resonator 1 does not have a second plateau present in the 1 to 100 photon range.
Furthermore, none of the resonators loss curves saturate at high powers.
One possible explanation is the renormalization of the critical exponent for the TLS bath due to TLS-TLS interactions, but this does not completely avoid a saturation effect~\cite{crowley_disentangling_2023,altoe_localization_2022,verjauw_investigation_2021}.
Another possibility is a large number of TLS baths with a wide distribution of critical photon numbers. 
Given these discrepancies with usual models, we do not attempt to fit to those models.

We conduct the resonator measurements in six different fridges (wiring diagrams in Appendix~\ref{app:fridge}).
The six systems, referred to as Meehl, NIST, SU, McMahon, McMahon2, and Fatemi were tested with a Ta resonator reference sample and showed good agreement in the extracted power dependent losses with published measurements (Fig.~\ref{fig:ta-calibration} in Appendix~\ref{app:resonator_measurement}).

The results of the substrate preparation series are summarized in Figure~\ref{fig:boxplots}(d).
We believe that the distributions across all the dies from the different substrate preparation methods are not sufficiently well-separated to draw a clear distinction.
As Nb resonators are susceptible to significant losses from the reformation of Nb oxide~\cite{zheng_nitrogen_2022,verjauw_investigation_2021,altoe_localization_2022,woods_determining_2019}, some of the variation in the data may be explained by the varying amount of Nb re-oxidation after the HF treatment.
In addition, temporal fluctuations of measured quality factor during the same cooldown for Nb on Si were found to have a relative standard deviation of 13\%~\cite{vallieres_loss_2024}, much smaller than what we observe cooldown-to-cooldown.

\subsection{Critical strip bath effects}
We find significant differences in device performance and surface chemical content when using the two different resist strip baths for the fabrication.
Figure~\ref{fig:boxplots}(d) shows the results from dies using the 1165 strip bath with varying durations of post-fabrication acid treatment, and Fig.~\ref{fig:boxplots_new} shows the results from dies using the 1165 strip bath with no post-fabrication acid treatment, and dies using the AZ strip bath without and with acid treatment.
For the recipe using 1165, a post-fabrication HF dip was necessary to achieve near-state-of-the-art (SOTA) performance.
Without the final HF dip, the low and high power losses were found to be about 15 times higher with median low power losses: $3.0\times10^{-5}$, $3.1\times10^{-5}$, and $3.2\times10^{-5}$ for the BOE, Anneal, and Thermal samples, respectively (compare the 1165 samples from Fig.~\ref{fig:boxplots_new} to the acid-treated samples in Fig.~\ref{fig:boxplots}(d)).
At high power, the loss reduces only modestly (less than a factor of two) for the 1165 samples with no acid treatment, indicating the dominant change is a source of power-independent loss.
Per Table~\ref{tab:contamination}, the Cl residue on the silicon surface (substrate-air interface) that results from the etch process is largely untouched by the strip bath but is removed by HF, as seen by XPS.
However, the AZ bath successfully removes chlorine and other residues as well as substantially reducing surface carbon content. 
This correlates with performance equaling to the levels seen with 1165 with the following HF dip, as shown in Fig.~\ref{fig:boxplots_new}. 
We tested with XPS two other NMP-based strip baths, and also found residual Cl on the Si surface (see Table~\ref{tab:contamination}) as well as elevated surface carbon content relative to AZ.

The Si oxide on the silicon surface across the different strip baths show no distinct correlations, with an average Si oxide of 10\% post-strip.

Within the Nb signal, Nb suboxides (NbO and NbO$_2$) account for around 10\% of the signal across all samples. The pentoxide is highest for AZ300T-treated films surfaces (60\%) and lowest for PGMEA-treated films (36.5\%). 
The higher pentoxide content in the AZ300T-treated sample decreased the LP performance as one would expect~\cite{verjauw_investigation_2021}, but even with the increased oxide thickness the AZ300T samples without post-HF treatment were comparable to the 1165 samples with post-HF treatment. 
Based on this observation, Cl residue likely has a comparable impact on the quality factor as the difference in the relative amounts of Nb oxide.
AZ300T samples with a \SI{1}{\minute} 2\% HF post-treatment have a similar level of oxide, as compared to the 1165 samples without post-HF treatment.
We highlight the importance of the resist stripping method used on our device performance.
See Appendix~\ref{app:strip_bath} for details on the parameters and the oxide fitting results.

The thicker surface oxide with the AZ300T samples motivated a return to attempting HF treatment following fabrication. 
With our usual \SI{60}{\second} treatment with 2\% HF, the losses lowered further to a median value $\delta_{\rm LP} = 0.83\times10^{-6}$ across three chips, as shown in Fig.~\ref{fig:boxplots_new}.
These three samples were consistently below the 25\% / 75\% loss window of samples treated with HF after the 1165 strip, or with AZ300T alone. 

We also examined samples with a \SI{20}{\minute} 10:1 post-BOE treatment, as an alternative to the post-HF treatment.
We found a median value of $0.48\times10^{-6}$ for $\delta_{\rm LP}$, as shown in Fig.~\ref{fig:boxplots_new}.
This performance exceeded that of the post-HF treatment, which could be indicative of hydrogen contamination in the HF-treated samples as previously reported~\cite{altoe_localization_2020,torres-castanedo_formation_2024}.
In addition, we re-measured these two samples after 4 and 6 weeks of exposure to ambient conditions without removal from the packages.
The aging process allows for both the SiO\textsubscript{2} and NbOx to regrow~\cite{verjauw_investigation_2021}.
We found that median after aging $\delta_{\rm LP}=1.27\times10^{-6}$, returning to the range around the median value of $1.52\times10^{-6}$ for samples without any post-acid treatment (see Fig.~\ref{fig:boxplots_new}).

\subsection{Additional film assessments}
\begin{figure*}
\centering
\includegraphics[width = 1.95\columnwidth]{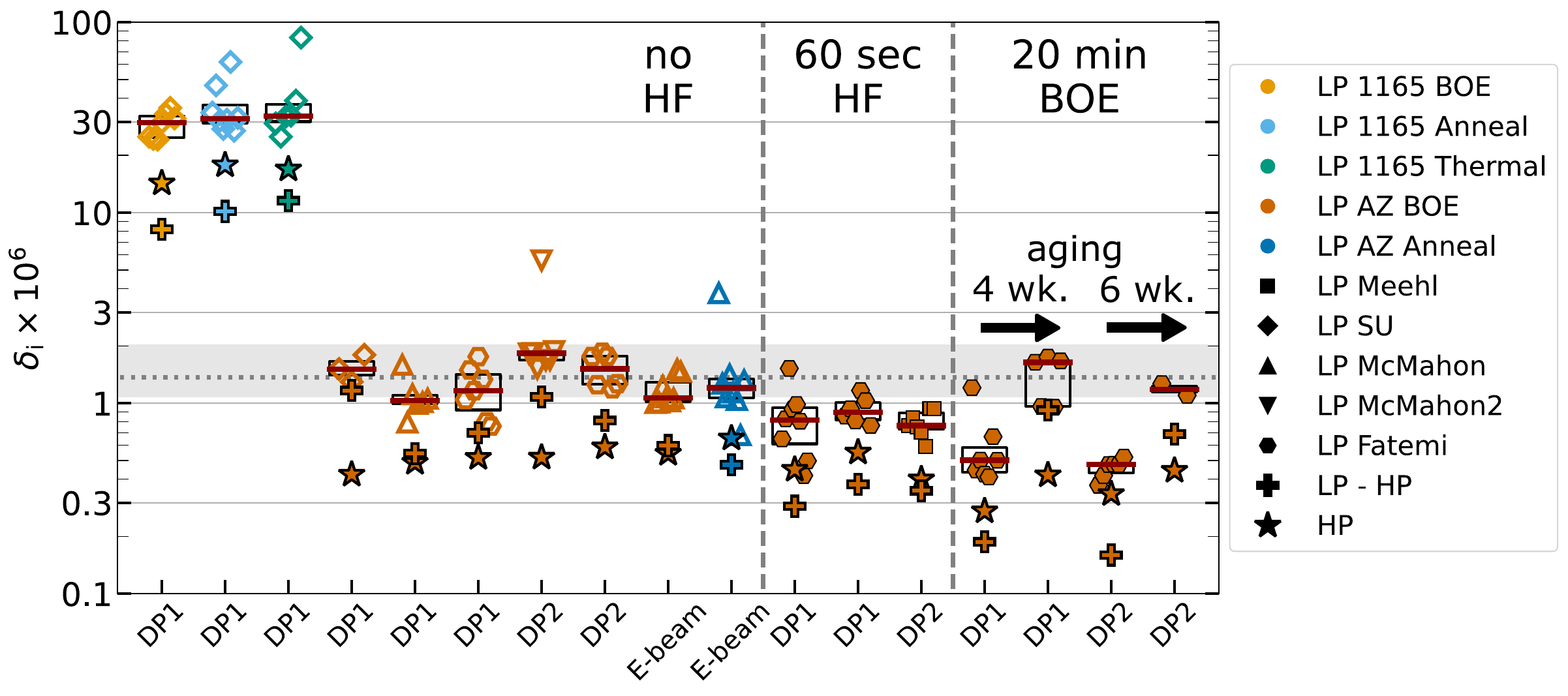}
\caption{
Performance of AZ300T resist strip. 
First three datasets use the 1165 strip bath with no post-fabrication acid treatment. 
The remainder all use the AZ procedure.
Films deposited with both DP1 and DP2 are shown.
E-beam labels Nb films deposited by electron-beam evaporation. 
The E-beam sample with blue markers had an additional \textit{in situ} anneal step for UHV oxide desorption.
The shaded band is the window of 25th to 75th percentile performance of the acid-treated resonators in Fig.~\ref{fig:boxplots}, with the dashed line being the median.
Vertical dashed line separates resonators without post-HF, with post-HF, and with post-BOE treatments.
The two post-BOE samples were aged for 4 and 6 weeks, as indicated by the arrows.}
\label{fig:boxplots_new}
\end{figure*}

With the updated AZ strip bath recipe (without HF), we have checked the performance of microwave resonators from niobium films deposited in two other configurations. 
First, Nb films were deposited in a chamber that are co-loaded with magnetic targets (DP2). 
Magnetic material depositions were conducted for weeks prior to the Nb depositions.
Prior to Nb film deposition, we accomplished the precautions described at the end of Sec.~\ref{sec:hygiene}. 
Fig.~\ref{fig:boxplots_new} shows that these films performed as well as any other Nb films with this fabrication recipe.
Second, Nb films were deposited by an electron beam evaporation tool on HF-treated Si(100) both with and without an \textit{in situ} anneal to accomplish UHV oxide desorption.
The combination of HF-treatment and UHV oxide desorption was found to significantly lower losses for aluminum resonators on silicon~\cite{earnest_substrate_2018}, but the resonators fabricated from our Nb films demonstrated similar losses to each other and to the sputtered films (Fig.~\ref{fig:boxplots_new}). 
This suggests substrate preparation and deposition method do not limit the loss; more detailed cross-sectional materials characterization and comparison will be accomplished in future work.

 \begin{table}
     \centering
        \begin{tabular}{|c|c|c|c|c|c|c|c|c|c|c|} \hline
         Surface & \multicolumn{6}{|c|}{Si}&\multicolumn{4}{|c|}{Nb}\\ \hline
         Element & C & Ca & Cl & N & Na &  F &C & Na & Si  & F\\ \hline\hline
        Pre-strip & 10.3 &  -& 1.4 & 0.8 &  -&  -&/ & /  & /  & / \\ \hline 
        AZ300T & 6.5 &  -&  -&  -&  -&  -&11.5 & 0.2 &   -&-\\\hline
        NMP & 8.9 &  -& 1.1 & 0.9 &  -&  -&18.0 &  -& 1.0  &-\\ \hline
        PGMEA & 6.8 & 0.4 & 1.0 &  -& 0.3 &  -&33.9 &  -&   -&-\\ \hline
        1165 & 7.9 &  -& 1.0 & 1.0 &  -&  -&19.3 &  -& 1.0  &-\\ \hline\hline
        AZ300T + 51D& 9.7& -& -& -& -& -& 12.4& 0.4& 3.4& -\\\hline
        AZ300T + HF& 13.1& -& -& -& -& 2.1& 13.0& -& 1.4& 2.0\\\hline
        1165 + 41D& 11.6 & -& 1.0 & 0.9& -& -& 22.0 & -& 1.8&-\\\hline
        1165 + HF&12.2  &  -&  -& 0.5 &  -&  2.8&17.6 &  -& 1.1  &1.7\\ \hline
        \end{tabular}
     \vspace*{3mm}
     \caption{
     Elements detected (in relative percentage points of the signal) by XPS on Nb and Si surfaces after our etch process (a mix of BCl\textsubscript{3}, Cl\textsubscript{2}, and Ar).  
     The ``Pre-strip" row is the measurement before stripping resist, so entries for Nb are labeled `/' because only the resist layer is seen. 
     Entries with a dash indicate the element was not detected.
     The HF dip check was on the same 1165-stripped and AZ300T-stripped chips, after 41 days (41D) and 51 days (51D), respectively. We see on the 1165 sample that Cl and N residues are persistent, and non-volatile carbon contamination has occurred.
     AZ300T appears to minimize carbonaceous residue and removes chlorine residue from the etch in the substrate-air interface.
     The residual F after HF dip desorbs on a timescale of hours.}
     \label{tab:contamination}
 \end{table}

\section{Discussion}\label{sec:discussion}
\begin{figure}
\centering
\includegraphics[width = 0.95\columnwidth]{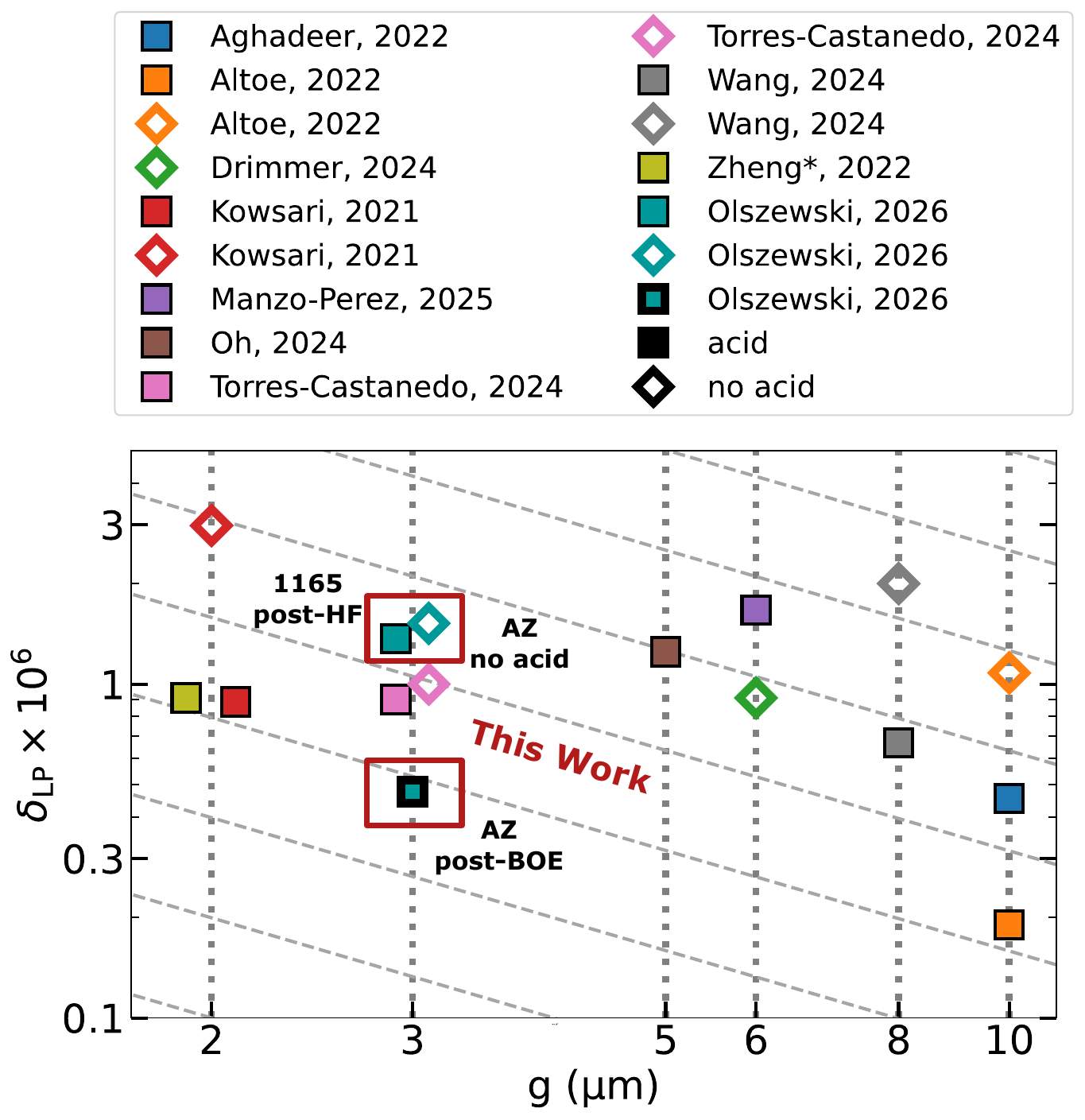}
\caption{
Scaling plot comparing our median losses to similar studies in the literature for Nb CPW resonators~\cite{mcrae_materials_2020,jones_grain_2023,alghadeer_surface_2023,altoe_localization_2022,oh_exploring_2024,torres-castanedo_formation_2024,zheng_nitrogen_2022,kowsari_fabrication_2021,wang_impact_2024,drimmer_effect_2024,manzo-perez_physical_2025}.
Solid markers indicate post-fabrication acid treatment and open markers indicate no acid treatment. 
The median LP losses for DP1, 1165 samples with post-HF treatment in our work were $\delta_{LP}=1.37\times 10^{-6}$ for BOE, Anneal, and Thermal treatments combined, marked with a teal square with a thin black outline. 
The median LP losses for AZ samples without post-HF in our work were $1.52\times10^{-6}$ for DP1 and DP2 combined, marked with an empty teal diamond. 
The median LP losses for AZ samples with post-BOE in our work were $0.48\times10^{-6}$ for DP1 and DP2 combined, marked with a teal square and thick black outline.
All of our samples correspond to $g=3$ $\mu$m; some points are offset horizontally for visibility.
Gray dashed lines approximate constant surface loss tangent with varied CPW resonator gap width~$g$.
Our losses are on par with state-of-the-art devices in the field, showing that our films are of comparable quality.
Similarly, the markers for Torres-Castanedo, 2024 are slight offset from $g=\SI{3}{\micro\meter}$, as well as Kowasari, 2023 and Zheng, 2022 from $g=\SI{2}{\micro\meter}$.
We remark that Zheng, 2022,~\cite{zheng_nitrogen_2022} used a rather deep trench of $\sim\SI{500}{\nano\meter}$ which likely significantly reduces the metal surface participation relative to the other works, and Torres-Castanedo, 2024,~\cite{torres-castanedo_formation_2024} uses a fluorine-based etch chemistry rather than chlorine.
}
\label{fig:scaling}
\end{figure}

Our results indicate the importance of solvent choices and sequences in Nb resonator fabrication. 
Although an HF dip is often used as a final step to improve resonator performance, the effectiveness of the HF dip appears to depend sensitively on the prepared quality of the surface.
With the HF dip, we find performance meeting the state-of-the-art, after accounting for the geometry~\cite{mcrae_materials_2020}, as shown in Fig.~\ref{fig:scaling}. 
However, this is only achieved with one resist strip bath (AZ300T) and not the other that we tested (1165). 
AZ300T includes, in addition to NMP, both a basic organic salt that helps to remove reacted metal compounds and a diol that helps to dissolve ionic elements (e.g. Cl). 
Either or both of which contributions are not provided by all strip baths, and we do find this bath results in the fewest measurable impurities on the surface.
This appears to set a better template for the HF dip.

Without the HF dip, we also find that the right choice of solvent bath can produce resonators of sufficient quality for readout and hybrid device purposes. 
Not all materials are compatible with acid-based processing, so this observation can motivate materials exploration for superconducting circuits.
Avoidance of HF-based processing can further mitigate degradation from hydride-formation in the Nb film~\cite{torres-castanedo_formation_2024},  and simplified fabrication processes can be helpful for reducing the number of points of failure. 
Absence of the HF dip can also be advantageous from a health and safety perspective, and fluorine-based nanofabrication has the potential to produce PFAS byproducts which will potentially be heavily regulated due to environmental and health concerns~\cite{sunderland_review_2019,cousins_high_2020,gluge_overview_2020,kwiatkowski_scientific_2020,ober_review_2022}.

Moreover, we achieved these state-of-the-art resonators despite depositing the Nb in a magnetically-contaminated chamber, including cases with magnetic material depositions in the same run of the tool. 
This shows that we were successfully able to use this tool as a dual-use system.
The characterization of our films was unable to detect magnetic impurities in the bulk of our Nb films or at the substrate-metal interface even after high-temperature processing within the chamber.
For the impurities present at the surface of the Nb, we are unsure of their origin -- we have observed their concentration can increase or decrease with acid treatment after fabrication, and the change in Fe concentration does not noticeably correlate with the performance of the resonators.
In the future, resonator-based electron-spin resonance measurements could be used to identify lower concentrations of paramagnetic impurities than we have been able to detect so far~\cite{jayaraman_loss_2024}.

We also tested three different methods of preparing substrates, which did not yield clearly distinct measurements of device quality.
This outcome is consistent with the metal-substrate interface being subdominant sources of low-power loss~\cite{woods_determining_2019,wenner_surface_2011} or the Nb-Si interface being largely unaffected by these differences in preparation. 
In order to further improve the quality of our devices we will consider other changes to our fabrication methods and materials stack synthesis of bilayer and alloy films to mitigate Nb oxide formation~\cite{bal_systematic_2023}.
In the future, we will also explore co-loading superconductor and magnet sputter targets to investigate heterostructures for, e.g., $\pi$-junctions~\cite{birge_ferromagnetic_2024,kim_superconducting_2024}.

We hope that this work will enable others in the community to innovate on superconducting materials for quantum information science based on existing equipment. 

\section*{Data and code availability}
All data generated and code used in this work are available at: 10.5281/zenodo.13857880.

\begin{acknowledgments}
\textit{Personal Acknowledgements}

\noindent
We acknowledge the extended technical help from AJA Int.~in refurbishing the tool. In particular we thank Wendell Sawyer for his expertise.
We thank Sridhar Prabhu, Alen Senanian, and Wayne Wang for help with resonator measurements and Peter McMahon for fridge space for some of the resonator measurements.
We thank Melissa Hines and Azriel Finsterer for early discussions and advice regarding silicon processing.
We are grateful to CNF staff members who have guided our fabrication process, in particular Chris Alpha, Garry Bordonaro, Jeremy Clark, Tom Pennell, and Aaron Windsor.
Thanks to Michael Vissers for wet processing assistance.
Thanks also to NIST ERB reviewers Miranda Thompson and Anthony McFadden for valuable feedback.

\textit{Funding and Facility Acknowledgements}

\noindent
This material is based upon work supported by the Air Force Office of Scientific Research under award number FA9550-23-1-0706. Any opinions, findings, and conclusions or recommendations expressed in this material are those of the author(s) and do not necessarily reflect the views of the United States Air Force.
M.O. was supported by the US National Science Foundation (DMR-2104268).
The contributions of McRae and Ramirez are funded by the Materials Characterization and Quantum Performance: Correlation and Causation (MQC) program by the Laboratory for Physical Sciences.
We thank the Kavli Institute at Cornell for funding towards improvement of the deposition chamber used in this work. 

The SIMS analysis in this paper was performed at the Metrology Services lab at the NY CREATES Albany NanoTech Complex. 
This work made use of the Meehl cryostat donated by David W. Meehl in memory of his father James R. Meehl and supported by the Cornell College of Engineering.
This work was performed in part at the Cornell NanoScale Facility (CNF), an NNCI member supported by NSF Grant NNCI-2025233.
This work made use of the Cornell Center for Materials Research (CCMR) shared instrument facilities.

Certain equipment, instruments, software, or materials are identified in this paper in order to specify the experimental procedure adequately. Such identification is not intended to imply recommendation or endorsement of any product or service by NIST, nor is it intended to imply that the materials or equipment identified are necessarily the best available for the purpose.

\end{acknowledgments}

\section*{Author Contributions}
M.O. accomplished the thin film synthesis, resonator fabrication, and materials characterizations at CCMR and CNF with the guidance of G.D.F., D.C.R., and V.F., and with the assistance of J.T.P., T.B., and H.L.
M.O. and J.T.P. resolved the changes in resonator performance with respect to the strip bath.
A.B. accomplished the SIMS measurements and analysis in coordination with M.O. and V.F.
R.P. synthesized the MBE Nb on Si under supervision of  C.J.K.R.  
T.B. accomplished the XPS measurements and analysis. 
M.O., J.T.P., H.L., J.L.R., and K.O. accomplished the resonator measurements and analysis with guidance from V.F., B.L.T.P., I.V.P., and C.R.H.M. 
M.O. wrote the manuscript with V.F. and C.R.H.M. 
All authors reviewed and provided feedback on the manuscript.

\section*{Competing Interests}

The authors declare no competing interests.

\clearpage
\onecolumngrid

\renewcommand{\thefigure}{S\arabic{figure}}
\renewcommand{\thetable}{S\arabic{table}}

\setcounter{figure}{0}
\setcounter{table}{0}
\appendix

\begin{titlepage}
  \centering
  \vskip 60pt
  \LARGE \textbf{Supplementary Material for ``Low-loss Nb on Si superconducting resonators from a dual-use spintronics deposition chamber and with acid-free post-processing''} \par 
  \par
  \vskip 2em
\end{titlepage}
  
\maketitle

\balancecolsandclearpage
\clearpage
\onecolumngrid

\balancecolsandclearpage
 
\linenumbers

\begin{table}
    \centering
    \begin{tabular}{c|c|c|c|c|c||c|c|c}
        \multicolumn{1}{c}{}&\multicolumn{1}{c}{} &\multicolumn{1}{c}{} &\multicolumn{1}{c}{} &\multicolumn{1}{c}{} & & \multicolumn{3}{c}{$\delta_{\rm LP}\times10^{6}$} \\
        Chamber & Prep & Strip & Acid & N\textsubscript{c} & N\textsubscript{p} & med & iqr & std \\
        \hline
        \hline
        DP1 & BOE & 1165 & N & 1 & 7 & 29.7 & 7.4 & 4.2 \\
        DP1 & Anneal & 1165 & N & 1 & 8 & 31.1 & 7.2 & 11.3\\
        DP1 & Thermal & 1165 & N & 1 & 6 & 32.1 & 7.1 & 19.6 \\
        DP1 & BOE & 1165 & HF & 4 & 29 & 1.56 & 2.35 & 2.53\\
        DP1 & Anneal & 1165 & HF & 5 & 31 & 1.57 & 0.93 & 2.53\\
        DP1 & Thermal & 1165 & HF & 5 & 33 & 1.20 & 0.52 & 0.70\\
        \hline
        DP1 & BOE & AZ & N & 5 & 30 & 1.45 & 0.65 & 0.47\\
        DP2 & BOE & AZ & N & 2 & 13 & 1.76 & 0.32 & 1.09\\
        DP1 & BOE & AZ & HF & 2 & 14 & 0.841 & 0.198 & 0.264\\
        DP2 & BOE & AZ & HF & 1 & 7 & 0.762 & 0.158 & 0.116\\
        DP1 & BOE & AZ & BOE & 1 & 7 & 0.501 & 0.150 & 0.262\\
        DP2 & BOE & AZ & BOE & 1 & 6 & 0.476 & 0.046 & 0.049\\
        DP1 & BOE & AZ & Aged & 1 & 5 & 1.64 & 0.71 & 0.36\\
        DP2 & BOE & AZ & Aged & 1 & 2 & 1.18 & 0.08 & 0.08\\
        \hline
        LPS & BOE & AZ & N & 1 & 8 & 1.06 & 0.27 & 0.19\\
        LPS & Anneal & AZ & N & 1 & 8 & 1.20 & 0.28 & 0.90
    \end{tabular}
    \vspace{3mm}
    \caption{Extracted LP losses for all devices measured.
    N\textsubscript{c} is the number of chips measured for the particular row, and N\textsubscript{p} is the total number of resonators measured.
    Each chip has eight resonators by design.
    The medians (med), interquartile ranges (iqr), and standard deviations (std) are included.
    }
    \label{tab:losses}
\end{table}

\section{Device Fabrication} \label{app:fabrication}
\subsection{Chamber preparation} \label{app:chamber}
We used two different procedure for preparing the deposition chamber.
In deposition procedure 1 (DP1), we only loaded superconducting targets.

\textbf{Deposition procedure 1 (DP1)}:
\begin{enumerate}[noitemsep]
    \item Bead-blast all detachable and non-fragile components (including gun assembly, substrate holder, quartz plate assembly, and shutters), then sonicate in acetone and isopropanol.
    \item Swap fragile components (screws, clips, and quartz plates) for a set used only for superconducting depositions.
    \item Scrub chamber with 3M Scotch-Brite.
    \item Install superconducting targets and pump down.
    \item Season chamber with Nb, sputtering with an open shutter and a high process gas pressure of 20-50 mTorr for about 30 \si{\min}. Then deposit Ti at \SI{250}{\watt} and 30 mTorr pressure until chamber base pressure is $10^{-9}$ Torr (about 30 \si{\min}).
    \item Individually season each sputter gun with the shutter closed in order to cover the shutter, chimney, and other local components with the sputter target material.
\end{enumerate}

For deposition procedure 2 (DP2) we deposited Nb films while co-loaded targets included permalloy (Ni$_{80}$Fe$_{20}$), cobalt (Co), and cobalt-iron-boron (Co$_{20}$Fe$_{60}$B$_{20}$).

\textbf{Deposition procedure 2 (DP2)}:
\begin{enumerate}[noitemsep]
    \item Follow DP1, except for cleaning the substrate holder and  swapping screw set, for which we used the ones normally dedicated for magnetic depositions.
    \item Season substrate holder with Nb for about 30 \si{\min}.
\end{enumerate}

\subsection{Substrate preparation}\label{app:substrate}

\begin{table}
    \centering
    \begin{tabular}{c|c|c|c}
        Preparation & BOE & Anneal & Thermal\\
        \hline
        T\textsubscript{c} (Before) & \SI{9.08}{\kelvin} & \SI{8.93}{\kelvin} & \SI{8.96}{\kelvin}\\
        RRR (Before) & 5.09 & 3.96 & 4.30\\
        RMS (Before) & \SI{250}{\pico\meter} & \SI{280}{\pico\meter} & \SI{240}{\pico\meter}\\
        T\textsubscript{c} (After) & \SI{9.05}{\kelvin} & \SI{8.75}{\kelvin} & \SI{8.74}{\kelvin}\\
        RRR (After) & 4.50 & 3.65 & 3.37\\
        T\textsubscript{c} (M-B) & \SI{8.04}{\kelvin} & \SI{8.12}{\kelvin} & \SI{8.25}{\kelvin}\\
        $\delta_{\rm LP}\times10^{6}$ (Median) & 1.56 & 1.57 & 1.20\\
    \end{tabular}
    \vspace*{3mm}
    \caption{Comparison of characterization measurements for Nb samples deposited using the three different substrate preparation methods, fabricated with the 1165 method. 
    `Before' refers to films measured right after deposition, and `after' denotes measurements done after resonator fabrication.
    As our fabrication procedure requires full 4" wafers, we were unable to measure the $T_{c}$ and RRR for all three substrates before the fabrication process.
    $T_{c}$ (M-B) fitted from Mattis-Bardeen fits from Eq.~\ref{eq:M-B}.
    }
    \label{tab:substrates}
\end{table}

\begin{figure}
\centering
\includegraphics[width = 0.95\columnwidth]{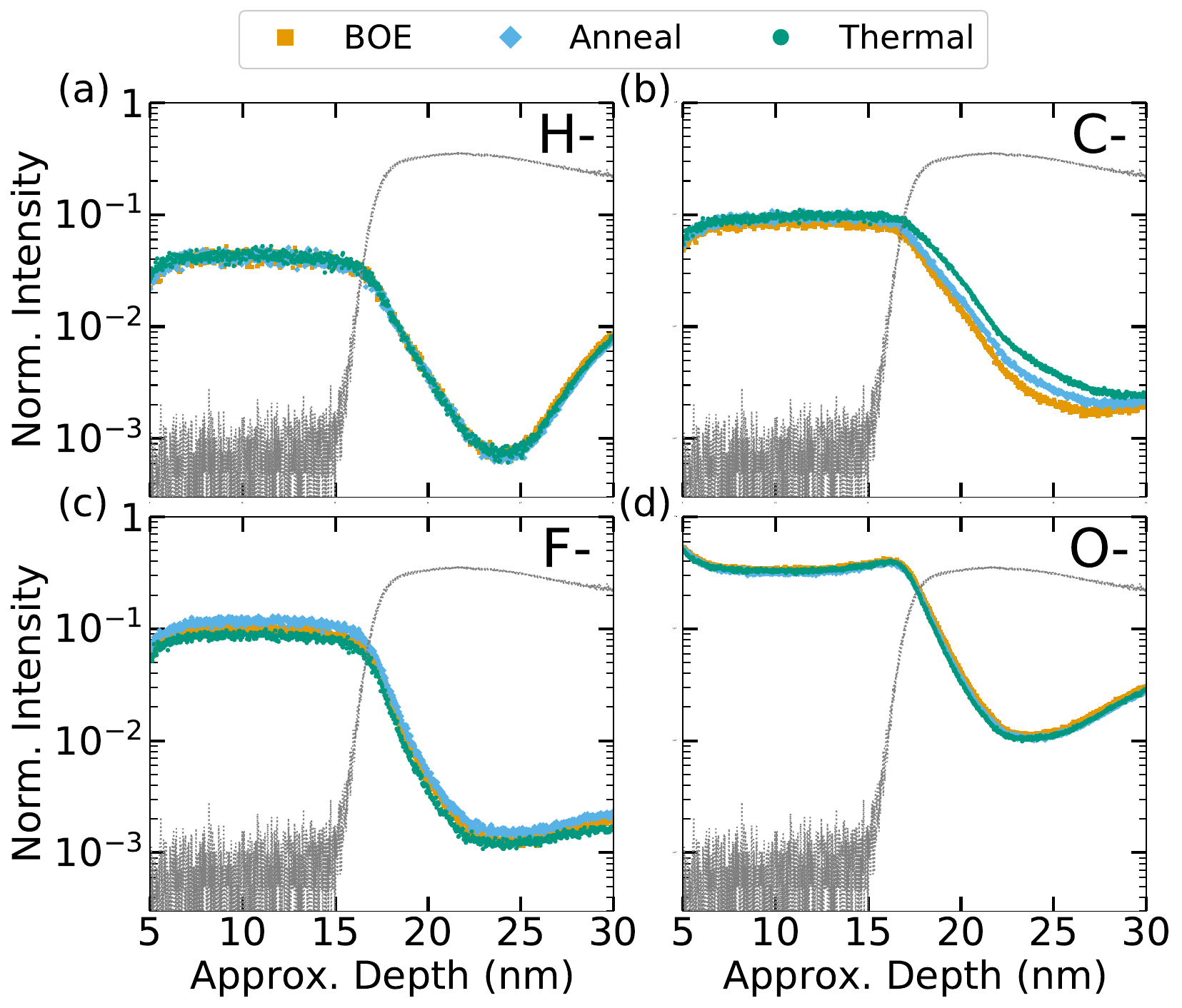}
\caption{SIMS profiles of (a) hydrogen, (b) carbon, (c) fluorine, and (d) oxygen contamination for the three different substrate preparation methods with the 1165 recipe, normalized to total secondary ion counts.
Dashed black lines plot the Si profile, indicating the Si substrate.
Analysis focuses on differences in the bulk of the Nb film and the substrate-metal interface.
In particular, we see qualitative differences in the amount of fluorine and carbon present across the three samples.
}
\label{fig:SIMS-Substrates} 
\end{figure}

The \textbf{BOE} preparation begins a two-step RCA Clean: SC-1 and SC-2. SC-1 is a 10-min soak in a 6:1:1 mixture of DI water, ammonium hydroxide, and hydrogen peroxide, respectively.
The clean is done at $70^{\circ}$C, followed by a 10-min rinse in DI water.
SC-2 is a 10-min soak at $70^{\circ}$C in a 6:1:1 mixture of DI water, hydrochloric acid, and hydrogen peroxide, respectively.
The clean is followed a 10-min water rinse and nitrogen gas dry.
Within one hour of completing the RCA clean the silicon oxide is removed by a 10:1 buffered oxide etch (BOE) soak for 1 minute, two 30-second deionized water rinses, and a blow dry with nitrogen.
The wafer is loaded into the deposition chamber load lock within 5 min of concluding the BOE etch process.
X-ray photoelectron spectroscopy (XPS) shows no significant oxide regrowth within 1 hour of the BOE etch, as shown in~\ref{fig:xps-Nb-survey}(b). 
We pump on the load lock for 1.5-2 hours, reaching pressures below $8\times10^{-7}$ Torr before transferring the wafer into the chamber for deposition.

The \textbf{Anneal} preparation includes all steps of the BOE preparation, followed by 
an in situ ``flash anneal", heating the wafer to $>\SI{700}{\celsius}$ for 3-4 \si{\minute}.
The wafer is then left to cool overnight (about 12 hours) before depositing the Nb film. The anneal process is a good test of the thermal robustness of the chamber hygiene methods. Additionally, annealing of silicon substrates before deposition has been shown to decrease aluminum resonator losses~\cite{earnest_substrate_2018,megrant_planar_2012,chayanun_characterization_2024}.

The \textbf{Thermal} preparation includes removal of the top layer of material from the wafer by a method adapted from~\cite{jakob_influence_1991}.
For the thermal oxide growth, we first ``cleaned" the furnace by running a 30 min, \SI{1000}{\celsius} dry oxidation with an additional flow of hydrochloric gas on an empty chamber.
The purpose of this is to remove ionic compound residue from the chamber.
After this, the chamber was loaded with substrates, which had just cleaned with the two-step RCA clean.
The dry thermal oxidation (without the use of HCl) was run for 6 hours resulting in about \SI{160}{\nano\meter} of silicon oxide~\cite{mcguire_semiconductor_1988}. 
The oxidation was followed by a 30 min nitrogen anneal at \SI{1000}{\celsius}, done in the same chamber.
The annealing step helps to create a uniform interface between the substrate and thermal silicon oxide.
This layer is then removed by a \SI{5}{\minute} 10:1 BOE soak, resulting in the removal of the top \SI{160}{\nano\meter} of silicon oxide, corresponding to roughly \SI{100}{\nano\meter} of the un-oxidized Si.
After this, the wafer undergoes the flash anneal and cooldown described as part of the Anneal preparation.

These substrate preparation methods were further compared using SIMS in Figure~\ref{fig:SIMS-Substrates}.
Here, we see no detectable differences in the hydrogen and oxygen contamination for the three samples.
For carbon, however, we observe that both the Anneal and Thermal samples had an increased presence of carbon which could be attributed to the overnight cool-down time in the chamber.
In the case of fluorine, we see that the Anneal sample had the highest presence, while BOE and Thermal were comparable.
Given the same variations in the overall quality of our devices, we cannot reach any firm conclusions about the substrates based solely on this SIMS data.

\subsection{Niobium sputtering deposition}
All Nb depositions were done at 2 mTorr and 70 sccm flow of ultra high-purity Ar gas from Airgas and a 3N5 purity, 2" diameter, and 1/8" thick target from AJA International Inc.
With a target-to-substrate distance of around 20 to 30 \si{\centi\meter}, these depositions had a rate of about \SI{2.4}{\angstrom\per\second} at \SI{325}{\watt}.
The substrate was rotated at 50 rpm during the depositions.
The base pressure varied between 1$\times 10^{-9}$ to 2$\times 10^{-9}$ Torr.

To develop the Nb deposition recipe, we compared the quality of various \SI{60}{\nano\meter} Nb films by measuring the RRR and $T_{c}$.
We optimized for the highest values of both RRR and $T_{c}$, by varying: target power, argon pressure, argon flow rate, substrate rotation, and substrate-to-target distance.
We found some correlation between these parameters and the value of RRR and $T_{c}$, noting the biggest change when varying the power and the substrate-to-target distance.
For the depositions reported in the main text, we maximized the deposition power and minimized the throw distance.

\subsection{Niobium e-beam deposition}
The e-beam deposited films were completed from a 99.95\% pure Nb starter source loaded into a 40cc UHV single pocket electron beam source.
The growth chamber base pressure is less than \SI{1e-10}{Torr} and the growth rate was \SI{0.13}{\angstrom\per\second}.
Prior to growth the 3-inch-diameter, float-zone-refined Si(001) wafer with resistivity \SI{20}{k\ohm\cm} was degreased with sequential baths of acetone, methanol, and isopropyl alcohol, and the native oxide was removed by a 5\% hydroflouric acid etch for \SI{120}{\second}, followed by a \SI{3}{\second} deionized water rinse.
The wafer was loaded into the HV load lock and heated to \SI{200}{\celsius} for \SI{4}{\hour} to desorb water before transferring into the growth chamber.
Niobium was deposited onto the substrate at \SI{25}{\celsius}.
The first film was grown on the wafer as transferred.
The second wafer was heated \textit{in-situ} to \SI{800}{\celsius} for \SI{15}{\minute} using a ramp rate of \SI{30}{\celsius\per\minute} and then cooled to \SI{25}{\celsius} for growth.
Reflection high energy electron diffraction was completed after growth showing rings of a polycrystalline film for the sample non-annealed substrate's sample, but a spotty/streaky pattern for the annealed substrate's sample.
Also, the sample with in situ annealed substrate expressed weak Nb(110) peaks in the symmetric $2\theta-\omega$ x-ray diffraction scan, while the sample without the annealed substrate did not express any Nb diffraction peaks.
Atomic force microscopy measured an rms roughness of \SI{1.2}{\nano\meter} for the Nb film with non-annealed substrate and \SI{0.96}{\nano\meter} for the annealed substrate using \SI{2}{\micro\meter} × \SI{2}{\micro\meter} scans.

\subsection{Device patterning}\label{app:dev-pattern}
We spun MP S1813 photoresist at 3000 rpm, then baked for \SI{1}{\minute} at \SI{90}{\celsius}. Our devices were patterned using a G-line stepper and developed in MP 321 MIF developer for \SI{1}{\minute}. After development, the sample was pre-etched with a \SI{60}{\second} low-power oxygen plasma in a RIE Oxford 80.
After loading into a Pan ICP-RIE Plasma-Therm 770 etcher, the sample is subjected to a light etch with a mixture of BCl\textsubscript{3}, Cl\textsubscript{2}, and Ar gas in a 2:30:5 sccm ratio with a RIE/ICP power of 26/800 W at 13 mTorr. This is followed by a primary etch with a mixture of BCl\textsubscript{3}, Cl\textsubscript{2}, and Ar gas in a 30:20:5 sccm ratio with a RIE/ICP power of 12/800 W at 7 mTorr.
The etch resulted in a roughly \SI{40}{\nano\meter} overetch into the silicon substrate.

For the samples treated with \textbf{MP 1165}, after the etch the sample was soaked in MP 1165 resist remover (1165) \SI{80}{\celsius} overnight (over 12 hours), followed by \SI{15}{\min} sonications in 1165, acetone, and isopropanol.
A \SI{10}{\min} oxygen plasma clean in the Oxford 80 was run after wet resist strip.
Prior to dicing the wafers, the wafer was once again coated with S1813 resist.
After dicing, the resist was once again stripped with a heated 1165 soak overnight, and sonication in: 1165, acetone, and isopropanol, followed by another \SI{10}{\min} oxygen plasma clean on the Oxford 80.

For the samples treated with \textbf{IMM AZ 300T}, after the etch the sample was soaked in heated AZ 300T resist stripper (AZ) at $80-90^{\circ}$C for 1 hour with a magnetic stir bar agitation, followed by \SI{10}{\min} sonications in isopropanol, and deionized (DI) water.
The process was finished with a rinse in a second water bath, and a second isopropanol bath.
Prior to dicing the wafers, the wafer was once again coated with S1813 resist.
After dicing, the resist was once again stripped with heated AZ stripper for 1 hour, followed by \SI{10}{\min} sonications in AZ, isopropanol, and DI water.
The process was finished with a rinse in a second water bath, and a second isopropanol bath.
Before packaging the samples was cleaned for \SI{60}{\sec} in a gentle oxygen plasma.

\section{Film Characterization}

\begin{figure}
\centering
\includegraphics[width = 0.95\columnwidth]{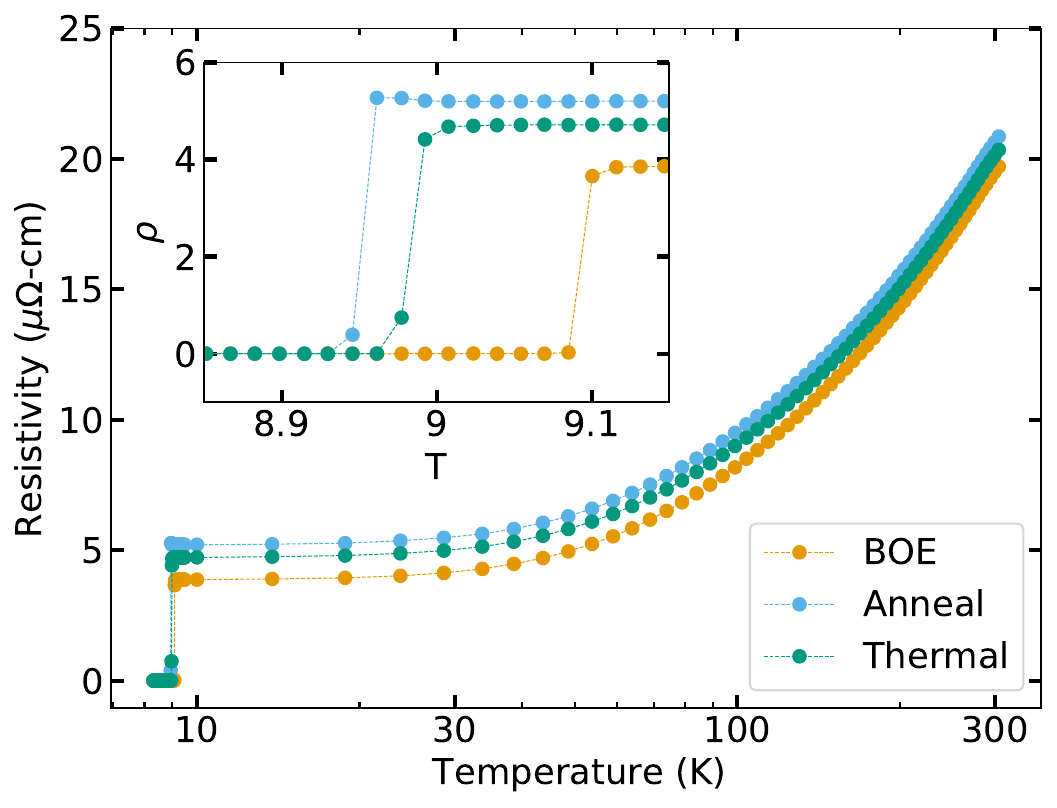}
\caption{Measurement of transition temperature for un-patterned \SI{60}{\nano\meter} films of Nb deposited on a BOE, Anneal, and Thermal substrates. Insert, shows a zoomed-in plot around the critical temperature of \SI{8.93}{\kelvin}, \SI{8.96}{\kelvin}, and \SI{9.08}{\kelvin}, respectively.}
\label{fig:transition} 
\end{figure}

\subsection{Film resistance in electronic transport}

\begin{figure*}
\centering
\includegraphics[width = 1.95\columnwidth]{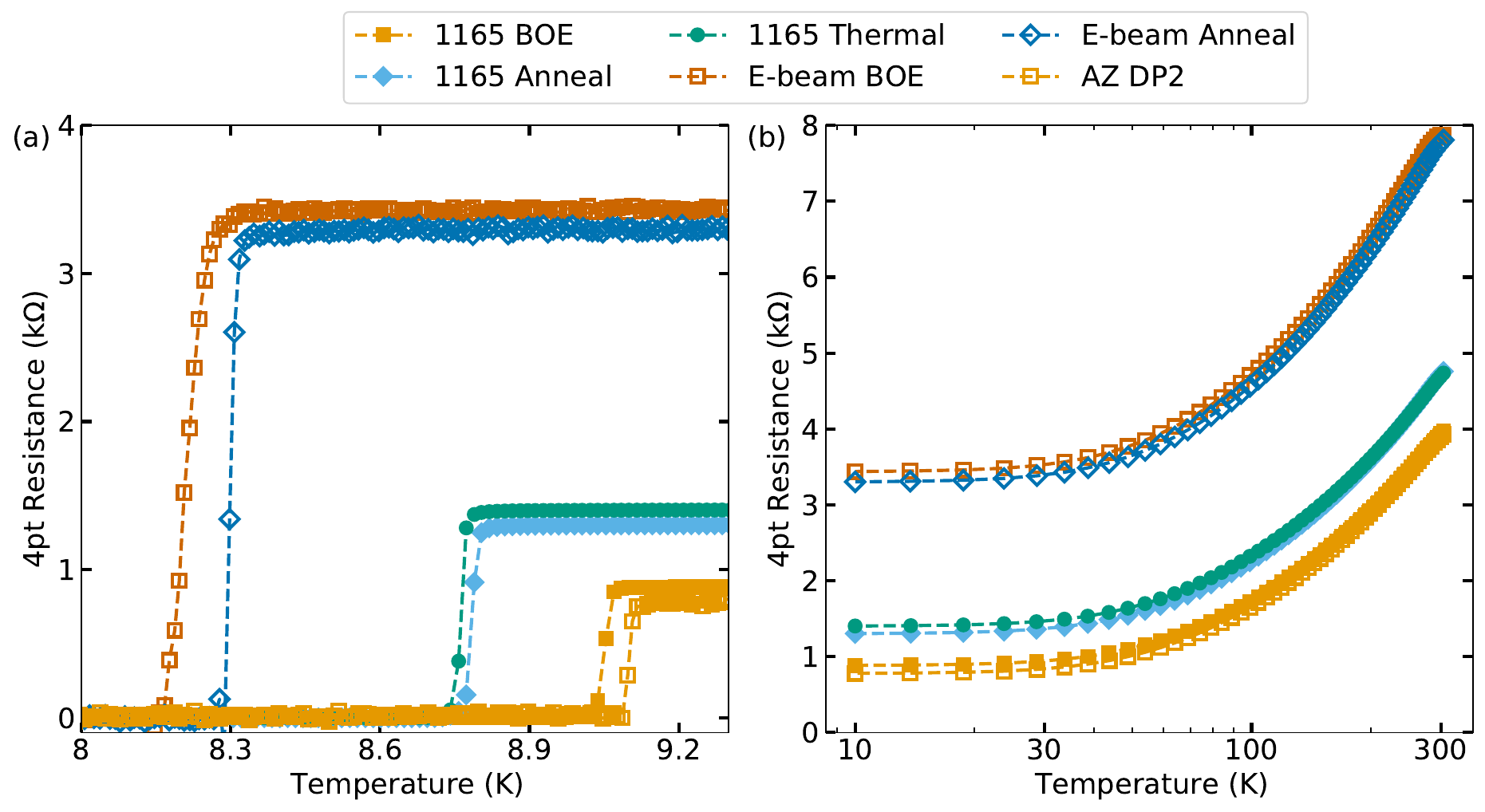}
\caption{Superconducting transition temperature for resonators with the various preparation methods after completion of the entire fabrication process, including the HF post-treatment for the filled shapes.
All resonators were fabricated with the same design.
(a) Shows the critical temperatures, measured to be \SI{9.05}{\kelvin}, \SI{8.75}{\kelvin}, and \SI{8.74}{\kelvin}, for BOE, Anneal, and Thermal samples, respectively, fabricated with DP1 and 1165.
While the $T_{c}$ for E-beam BOE, E-beam Anneal, and DP2 samples are \SI{8.15}{\kelvin}, \SI{8.27}{\kelvin}, and \SI{9.07}{\kelvin}, respectively, fabricated with AZ.
(b) Plots the temperature of the various patterned resonators during cool down.
The RRR for BOE, Anneal, and Thermal are 4.49,3.65, and 3.37, respectively. For e-beam BOE, e-beam Anneal, and DP2 they are 2.29, 2.36, and 5.03.}
\label{fig:tc_post} 
\end{figure*}

We characterized un-patterned films by measuring the residual-resistance ratio (RRR) between room temperature and \SI{10}{\kelvin} and the superconducting critical temperature ($T_{c}$).
In Figure~\ref{fig:transition}, we show a sample plot of the critical temperature for un-patterned \SI{60}{\nano\meter} Nb films deposited on various substrate preparations.

In comparison, the patterned resonator devices, had lower values for the $T_{c}$, shown in Figure~\ref{fig:tc_post}.
Notably, the un-annealed sample (BOE) has a higher $T_{c}$ than the two wafers that have been annealed (Anneal and Thermal).
We do not see any notable differences between the BOE and the DP2 sample.
The e-beam deposited samples did have a significantly lower $T_{c}$, as compared to the sputter counterparts.
Of note is that the e-beam samples had a reduced $T_{c}$ and RRR.

\subsection{SIMS measurements}

\begin{figure}
\centering
\includegraphics[width = 0.95\columnwidth]{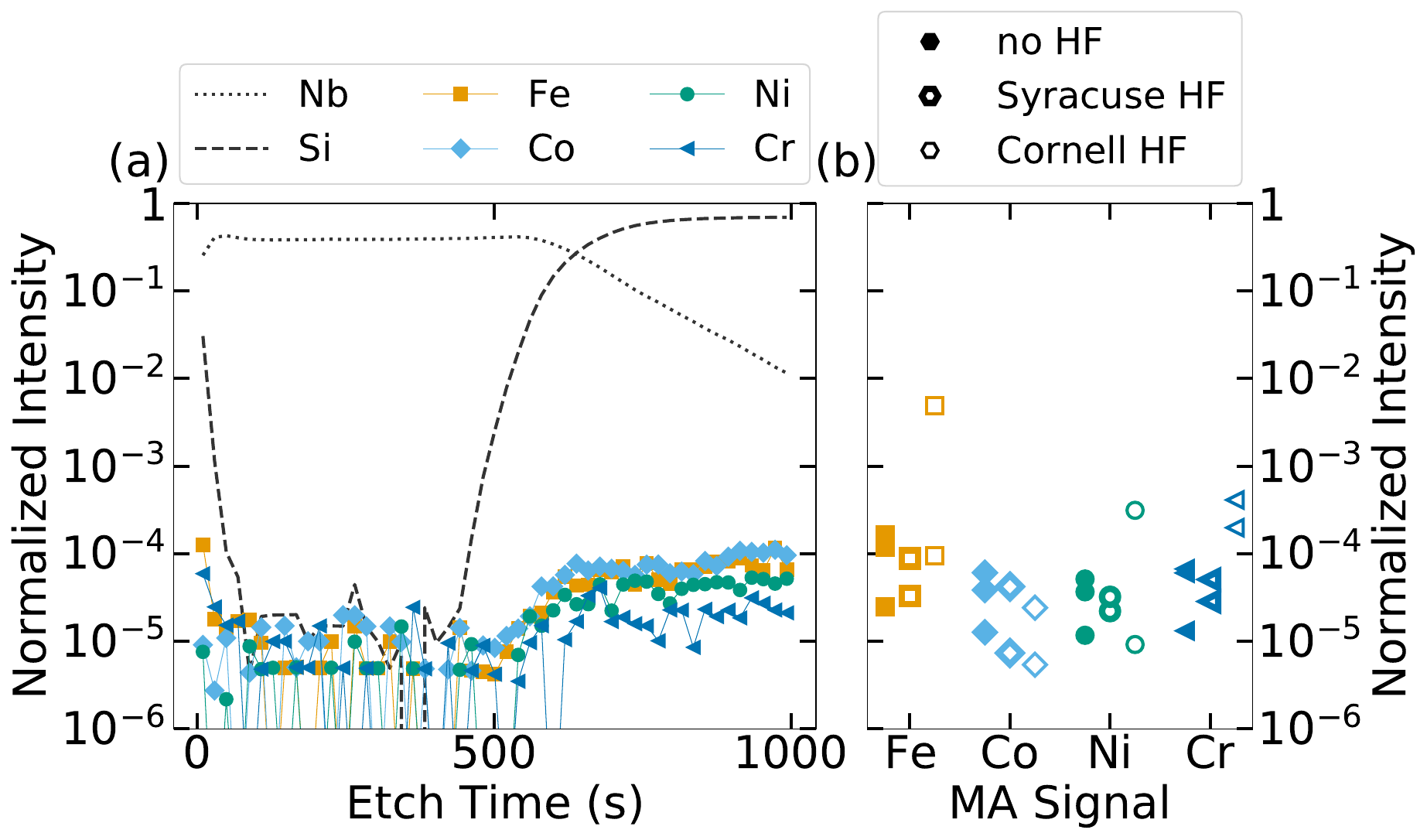}
\caption{(a) Depth dependent SIMS measurement of contamination of various magnetic elements. A sample showing metal-air interface magnetic presence is selected. (b) Subsequent measurements of multiple samples with treatment in 2\% HF for 60 seconds done at various locations.}
\label{fig:sims_check_append} 
\end{figure}

TOF-SIMS analysis was collected using a TOF-SIMS V (ION-TOF USA Inc, Chestnut Ridge, NY), equipped with a 25 keV Bismuth liquid metal ion gun (LMIG), and low energy Cesium and Oxygen ion sources.
The SIMS spectra were acquired with high mass resolution using Bi1+ in ‘Spectrometry’ setting, previously called high current bunched (HCBU). The Bi beam had a pulsed current of 1.2 to \SI{1.6}{\pico\ampere}, depending on TOF cycle time selected.

Dynamic SIMS data consisted of dual-beam instrument operation, by alternating stages of analysis (Bi+ primary ions) and material removal (O2+ or Cs+ sputter ions), until reaching substrate. Depth profiles for magnetic (metal) impurities (Fe, Co, Ni, Cr) were conducted with 1.0 keV O sputter beam (\SI{300}{\nano\ampere} current), for positively charged secondary ions. Positive polarity depth profiles consisted of \SI{300}{\micro\meter\squared} analysis area (raster size, or field of view), centered inside \SI{900}{\micro\meter\squared} erosion crater. Profiles for non-metal impurities (H, C, O, F) were performed with 0.5 keV Cs sputter beam (\SI{30}{\nano\ampere} current), for negatively charged secondary ions. Negative polarity depth profiles were comprised of \SI{150}{\micro\meter\squared} analysis area, centered inside \SI{600}{\micro\meter\squared} sputter crater. In both positive and negative experiments, data was collected in interlaced mode, with electron flood gun (for charge compensation), and without external reference materials (for concentration or depth quantification).
The raw intensity of impurity ions was normalized to the total secondary ion counts in each profile.

For our MP 1165 device fabrication procedures, we have found that a post-fabrication hydrofluoric acid (HF) treatment is crucial for obtaining high quality factors~\cite{verjauw_investigation_2021}. 
The Fe surface contamination can be reduced with a 60-second 2\% HF soak, but this is not always the case, as shown in~\ref{fig:sims_magnet}(b).
Samples treated at different sites prior to measurement (Syracuse University and Cornell University) showed different trends in the Fe contamination but similar quality factors. 
Therefore, we hypothesize that the remaining Fe contamination on the surface of the sample is not a limiting source of loss.
Besides removing surface Fe contamination from our films, the HF post-treatment also removes a residual amount of re-deposited Nb found in regions where Nb was meant to be etched away.

\subsection{XPS measurements}\label{app:film-xps}

\begin{figure}
\centering
\includegraphics[width = 0.95\columnwidth]{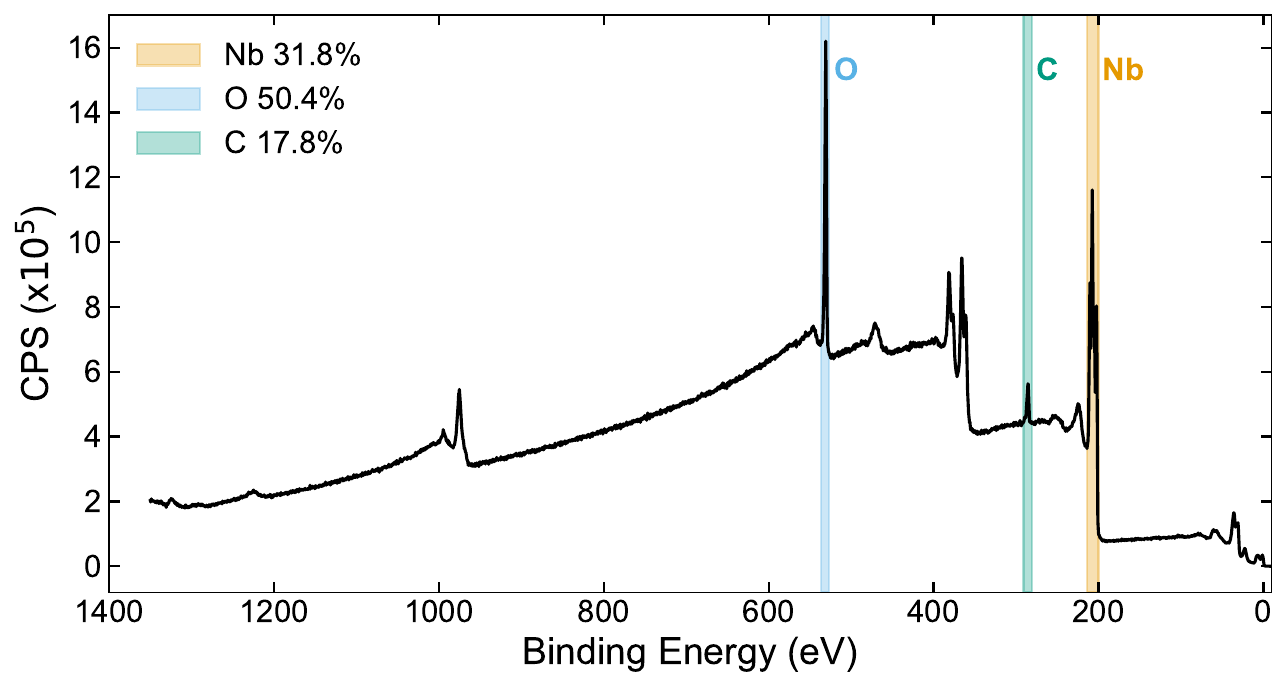}
\caption{XPS survey scan of Nb surface after deposition. Highlighted are the only detected elements: Nb, O, and C.
}
\label{fig:xps-Nb-survey} 
\end{figure}

\begin{figure}
\centering
\includegraphics[width = 0.95\columnwidth]{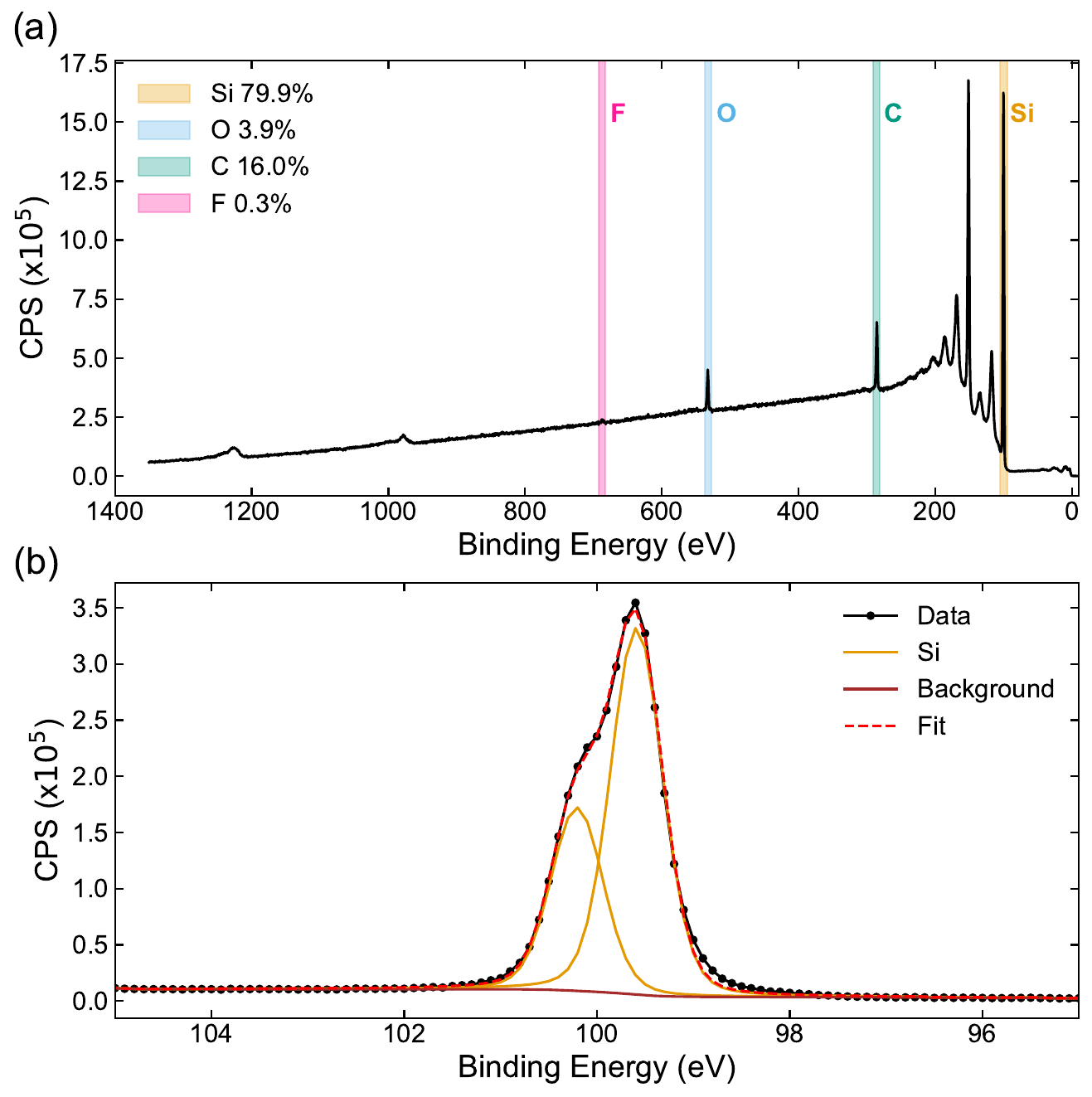}
\caption{(a) XPS survey scan of a Si surface 1 hour after a 1 min BOE dip. Highlighted are the only detected elements: Si, O, C, and F. The F signal is attributed to F-termination on the Si surface. The O signal is likely from adsorbed oxygen on the surface. (b) Si 2p core level XPS spectra. There is no evidence of any Si oxide formation on the surface.}
\label{fig:xps-Si-Scan} 
\end{figure}

All XPS measurements were conducted in a Thermo Scientific Nexsa G2 Surface Analysis system with a chamber pressure around $1.9\times10^{-8}$\,Torr. 
The system uses a monochromated Al K$\alpha$ X-ray source and the Fermi level is calibrated using a silver standard. 
The survey scan was collected using a 400 micron spot size as counts per second (CPS) with a 0.4 eV energy resolution. 
Fitting was done with a Shirley background in CasaXPS (v2.3.25)~\cite{fairley_systematic_2021} using Thermo Scientific provided sensitivity factors calibrated for the Nexsa.  

From the XPS survey scan of the Nb surface, Figure~\ref{fig:xps-Nb-survey}, we see no signs of any elements besides Nb, C, and O within the detection limit of XPS. 
This indicates that there are no major contaminations from the magnetically contaminated chamber. 

We also examined the surface of Si 1 hour after a 1 min BOE dip to ensure that no oxide had regrown on the surface. 
The survey scan, Figure~\ref{fig:xps-Si-Scan}(a), shows a minimal oxygen signal, likely from the adsorbed oxygen on the surface. 
The fluorine signal is attributed to F-termination of the Si and some remaining degassing residues.

The Si 2p core level spectrum is shown in Figure~\ref{fig:xps-Si-Scan}(b). 
The spectrum was collected with a 0.1 eV energy resolution, and peak fitting was done with a symmetric Voigt-like lineshape LA(50). 
The spin-orbit split peaks were constrained to have identical FWHM and a peak area ratio of Si2p$_{1/2}$:Si2p$_{3/2}$ of 1:2. 
The Si2p$_{3/2}$ had a binding energy of 99.58 eV and a spin-orbit split of 0.61 eV. 
There is no evidence of oxide formation on the Si surface, which nominally has a peak around 103.5 eV.
Any F-terminated Si would be too weak to detect as the relatively low concentration of F (0.26\%) implies a low SiF$_x$ concentration.

\subsection{AFM measurements}\label{app:afm}

\begin{figure}
\centering
\includegraphics[width = 0.95\columnwidth]{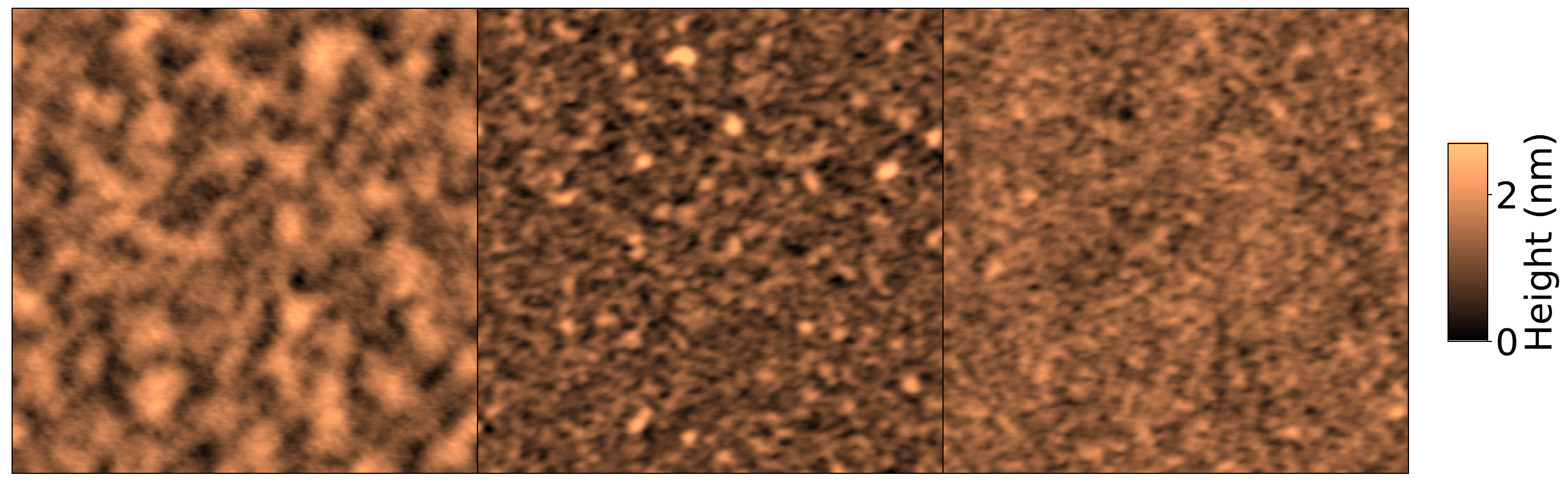}
\caption{From left to right, AFM surface roughness measurements for Nb films for the three different samples: BOE, Anneal, and Thermal, presented in the three columns, respectively. The scan area is \SI{2}{\micro\meter}~$\times$~\SI{2}{\micro\meter}.}
\label{fig:afm} 
\end{figure}

Using atomic force microscopy (AFM) we measured the surface roughness of Nb films deposited on wafers treated with the three different methods.
Changing the substrate and the preparation has been shown to alter the surface structure of both Al and Nb films.
In our measurements, shown in Figure~\ref{fig:afm}, we do not observe any significant qualitative or quantitative differences in the three films.
The film roughness, as report in Table~\ref{tab:substrates}, remains low ($<\SI{300}{\pico\meter}$) for un-patterned films.
The roughness for patterned films increases significantly to about 3 nm to 5 nm.

\section{Strip baths}\label{app:strip_bath}

\subsection{Sample preparation}

The sample preparation follows similar processes to the ones detailed in Appendix~\ref{app:fabrication}. 
The substrate was RCA cleaned prior to deposition. \SI{60}{\nano\meter} Nb was deposited at the Cornell Nanoscale Facility by DC sputtering with an AJA International, Inc., sputtering chamber. 
Devices were then patterned using the Plasma-Therm 770 etcher as detailed in App.~\ref{app:dev-pattern} until the photoresist strip process. 
The wafer was then cleaved into squares for the strip bath comparison test.

Four strip baths were compared alongside the non-stripped sample. 
The \textbf{AZ 300T} sample processed identically to the AZ 300T process done for the measured resonators. 
The same process was applied for the \textbf{1165} and the \textbf{NMP} baths by switching AZ 300T for the respective strippers. 
PGMEA cannot be heated to 70\textdegree C, and thus the \textbf{PGMEA} sample was sonicated in PGMEA for 15 minutes before following the same sonication and rinse process as the other samples.

The HF dip of the 1165 and the AZ300T samples took place one month after the initial measurement. 
To ensure consistency in measurement, the sample was remeasured prior to the HF dip (\textbf{1165 + 41D} \& \textbf{AZ300T + 51D}). 
The sample was then dipped in 2\%HF for 60s and measured (\textbf{1165 + HF} \& \textbf{AZ300T + HF}).

\subsection{Strip bath XPS measurements}\label{app:xps}

Each square cleaved for the measurement contained four resonator dies. 
The exposure pattern has the dies separated by a gap greater than 400 microns. 
Samples were measured in the same XPS system as in App.~\ref{app:film-xps}.
The gap provides the space necessary for the XPS to measure the Si surface. Each sample was measured at two spots: one was on the surface of the Nb on the resonator chip, and the second was at the center of the intersection of the gaps between the dies.

Measurements were done on the Nb surface with a 400 micron spot size and the flood gun off. 
Survey scans were used to determine contamination present on the surface. Detailed scan of the Nb 3d region with 0.1 eV energy resolution was used for the oxide fitting. 
Core level fitting was done using asymmetric Voigt-like lineshapes for the metallic Nb and NbO, which are conducting, while a Gaussian-Lorentzian mix GL(50) lineshape was used for the NbO\textsubscript{2} and Nb\textsubscript{2}O\textsubscript{5}~\cite{noauthor_x-ray_nodate}.
The spin-orbit splitting for Nb was set to 2.75 eV, with the peak area ratio Nb3d$_{3/2}$:Nb3d$_{5/2}$ of 2:3 and full width at half maximum restricted to being identical for spin-orbit split peaks as well as for the suboxides and oxide. 
See Table~\ref{tab:xps_Nb_oxide} for the curve fitting results of the Nb3d region.
\begin{table*}
     \centering
        \begin{tabular}{c|c|c|c|c|c|c|c|c} \hline

         Peak & \multicolumn{2}{|c|}{Nb} &\multicolumn{2}{|c|}{NbO}& \multicolumn{2}{|c|}{NbO\textsubscript{2}} &   \multicolumn{2}{|c}{Nb\textsubscript{2}O\textsubscript{5}}\\ \hline
         Strip  &  d$_{5/2}$ B.E. (eV)& At\% & d$_{5/2}$B.E.(eV)& At\% & d$_{5/2}$B.E.(eV) & At\% & d$_{5/2}$B.E.(eV) & At\%\\\hline\hline
         Pre-strip & -&  -& -& -&-& -& -&-\\ \hline 
         AZ300T & 202.33 eV&  30.3\%&  203.73 eV&  4.6\%&  206.36 eV& 5.5\%& 207.81 eV&59.7\%\\\hline
         NMP & 202.38 eV&  44.1\%& 203.68 eV& 7.4\%&  206.28 eV&  4.2\%& 207.78 eV&44.3\%\\ \hline
         PGMEA & 202.36 eV& 52.3\%& 203.74 eV&  7.7\%& 206.26 eV&  3.4\%& 207.68 eV&36.6\%\\ \hline
         1165 & 202.36 eV&  46.0\%& 203.69 eV& 7.6\%&  206.27 eV&  3.9\%& 207.76 eV&42.5\%\\ \hline \hline
         AZ300T + 51D& 202.20 eV& 29.1\%& 203.69 eV& 4.3\%& 206.19 eV& 4.9\%& 207.63 eV& 61.7\%\\ \hline
         AZ300T + HF& 202.18 eV& 59.8\%& 203.48 eV& 7.5\%& 206.1 eV& 4.5\%& 207.54 eV& 28.2\%\\ \hline
         1165 + 41D& 202.18 eV& 37.6\%& 203.59 eV& 5.9\%& 206.1 eV& 4.8\%& 207.53 eV&51.8\%\\\hline
         1165 + HF&202.16 eV&  62.1\%&  203.56 eV& 7.8\%&206.1 eV&  3.6\%& 207.51 eV&26.5\%\\ \hline

        \end{tabular}
    \vspace*{3mm}
     \caption{Nb 3d core level peak fitting results on the Nb surface of the resonator (metal-air interface).
     Atomic percentages are rounded to the nearest tenth. The Nb binding energy of the 3d 5/2 orbital is given, with the 3d 3/2 orbital fixed at 2.75 eV higher.
     There is no pre-strip data due to the thick photoresist layer on the Nb.
     }
     \label{tab:xps_Nb_oxide}
 \end{table*}

The measurement of the Si surface was done with a 100 micron spot size (to be sufficiently far away to not accidentally measure the Nb) and the flood gun off, with survey scans being used for contamination determination. 
Similar to the Nb 3d scan, the Si 2p scan had an energy resolution of 0.1 eV. 
The Si peak was fit using a spin-orbit split Voigt-like lineshape with a 1:2 ratio and identical FWHM. 
The Si oxide and organic Si were fitted using individual GL(50) lineshapes and constrained to have identical FWHM. 
The organic Si peak was allowed to vary between 101 and 102.5 eV, while the SiO\textsubscript{2} peak was constrained to be above 103 eV. See Table~\ref{tab:xps_Si_oxide} for the core level fitting results.
\begin{table*}
     \centering
        \begin{tabular}{c|c|c|c|c|c|c} \hline

         Peak & \multicolumn{2}{|c|}{Si} &\multicolumn{2}{|c|}{Organic Si}& \multicolumn{2}{|c}{SiO\textsubscript{2}}  \\ \hline
         Strip  &  p$_{3/2}$ B.E. (eV)& At\% & B.E.(eV)& At\% & B.E.(eV) & At\%\\\hline\hline
         Pre-strip & 99.03 eV&  89.5\%& 101.6 eV& 5.8\%&103 eV& 4.7\%\\ \hline 
         AZ300T & 99.29 eV&  90.3\%&  101.57 eV&  2.9\%&  103.24 eV& 6.8\%\\\hline
         NMP & 99.27 eV&  85.7\%& 101.59 eV& 2.4\%&  103.09 eV&  11.8\%\\ \hline
         PGMEA & 99.26 eV& 87.5\%& 101.73 eV&  3.6\%& 103.18 eV&  9\%\\ \hline
         1165 & 99.19 eV&  85.5\%& 101.27 eV& 2.5\%&  103.07 eV&  12.0\%\\ \hline \hline
         AZ300T + 51D& 99.32 eV& 85.4\%& 101.81 eV& 3.5\%& 103.28 eV& 11.1\%\\ \hline
         AZ300T + HF& 99.41 eV& 100\%& - & - & - & -  \\ \hline
         1165 + 41D& 99.15 eV& 82.5\%& 101.89 eV& 5.2\%& 103.2 eV& 12.4\%\\\hline
         1165 + HF&99.38 eV&  95.6\%&  101.45 eV& 1.8\%&102.95 eV&  2.6\%\\ \hline

        \end{tabular}
        \vspace*{3mm}
     \caption{Si 2p core level peak fitting results on the Si substrate of the resonator (substrate-air interface). Atomic percentages are rounded to the nearest tenth. The Si binding energy of the 2p 3/2 orbital is given, with the 2p 1/2 orbital fixed at 0.6 eV higher. No spin-orbital splitting was distinguished for the oxide or the organic peaks. The HF-dipped AZ300T sample did not have any oxide or carbonaceous Si signal.}
     \label{tab:xps_Si_oxide}
 \end{table*}

The variation in the binding energies of the organic Si is attributed to the variation in the bonding of Si to different organics. 
It is also likely that the Si could be bonded to the other contaminants (such as Cl or N in 1165, etc), which fall in the 101-102 eV range nominally~\cite{noauthor_x-ray_nodate}. 
The low signal prevents fitting of those elements to determine the exact compounds.

\section{Resonator Measurement} \label{app:resonator_measurement}

\subsection{Fridge setups}
\label{app:fridge}

\begin{table*}
    \centering
    \begin{tabular}{c|c|c|c|c|c||c|c}
        \begin{tabular}{@{}c@{}}Measurement\\Setup \end{tabular} & \begin{tabular}{@{}c@{}}Meehl\\ \SI{60}{\second} \end{tabular}  & NIST & \begin{tabular}{@{}c@{}}SU\\ \SI{18}{\second} \end{tabular} & \begin{tabular}{@{}c@{}}SU\\ \SI{60}{\second} \end{tabular} & \begin{tabular}{@{}c@{}}Fatemi\\ \SI{60}{\second} \end{tabular} & McMahon & McMahon2\\
        \hline
        HF Concentration & 2 \% & 2 \% & 2 \% & 2 \% & 2 \% & N/A & N/A \\
        HF Time & \SI{60}{\second} & \SI{60}{\second} & \SI{18}{\second} & \SI{60}{\second} & \SI{60}{\second} & N/A & N/A \\
        Reoxidation & \SI{3}{\hour} &  \SI{48}{\hour} &  \SI{29}{\hour} &  \SI{29}{\hour} &  \SI{3}{\hour} & N/A & N/A\\
        Base Temperature &  \SI{17}{\milli\kelvin} & \SI{11}{\milli\kelvin} & \SI{100}{\milli\kelvin} & \SI{100}{\milli\kelvin} & \SI{10}{\milli\kelvin} & \SI{48}{\milli\kelvin} & \SI{19}{\milli\kelvin}\\
        
    \end{tabular}
    \vspace*{3mm}
    \caption{Summary of differences in sample measurements for the different cryogenic systems, shown in Figure~\ref{fig:fridge}.}
    \label{tab:fridges}
\end{table*}

\begin{figure*}
\centering
\includegraphics[width = 1.95\columnwidth]{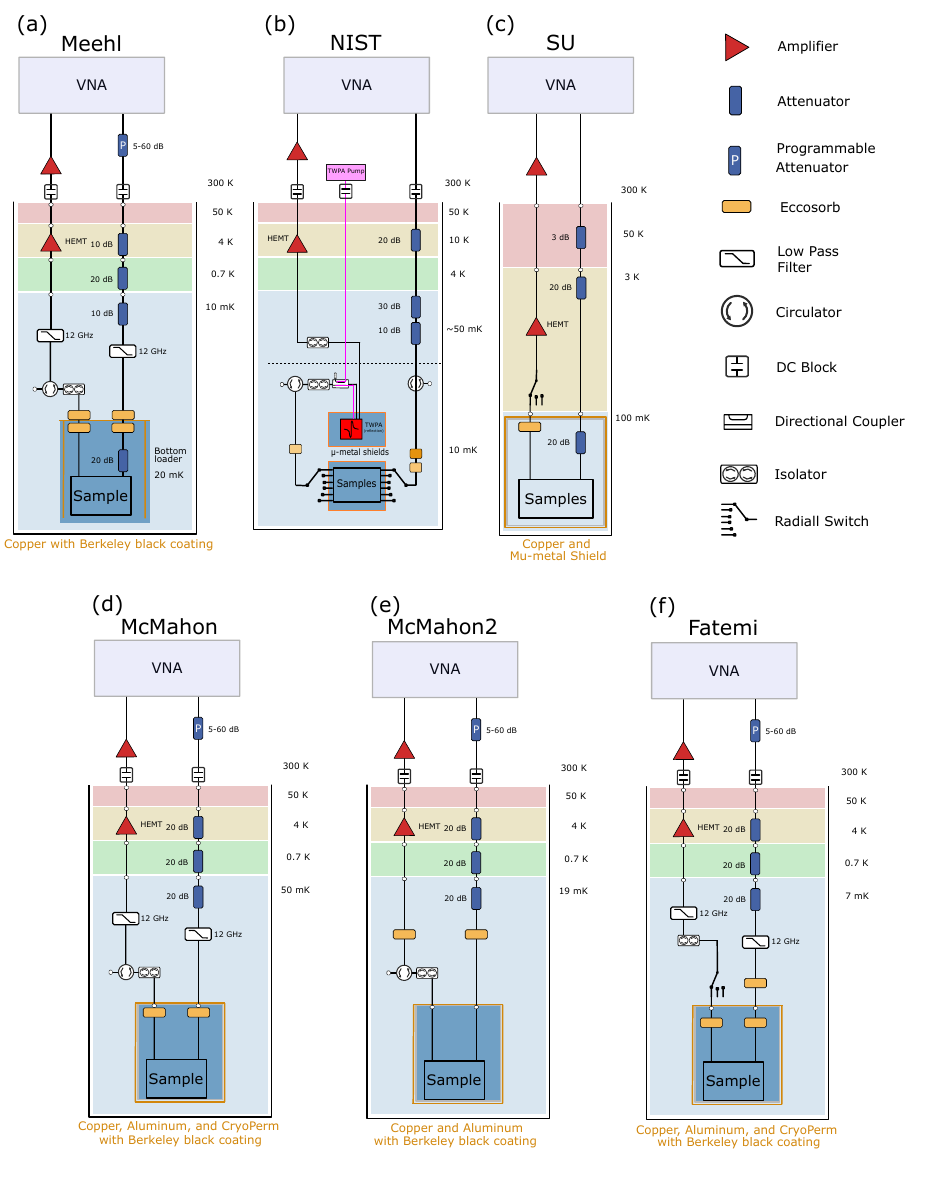}
\caption{Fridge diagrams for (a) Meehl, (b) NIST, (c) SU, (d) McMahon, (e) McMahon2, and (f) Fatemi measurement cryostats.}
\label{fig:fridge} 
\end{figure*}

All resonator measurements were conducted in five millikelvin refrigerators, labeled \textbf{Meehl}, \textbf{NIST}, \textbf{SU}, \textbf{McMahon}, \textbf{McMahon2}, and \textbf{Fatemi}.

The \textbf{Meehl} fridge is a Bluefors LD dilution refrigerator with bottom loader and a base temperature of \SI{17}{\milli\kelvin}.
As described in Figure~\ref{fig:fridge}(a), the input lines have \SI{60}{\deci\bel} attenuation from attenuators and \SI{7}{\decibel} attenuation from cabling inside the cryostat.
Copper Mountain M5180 VNA of 0 to \SI{-50}{\deci\bel m} dynamic range is connected to one programmable attenuator with tunability between 0 to \SI{60}{\deci\bel} attenuation at room temperature; the output lines host one high electron mobility transitor (HEMT) amplifier of \SI{37}{\deci\bel} gain, and one amplifier at room temperature of \SI{38}{\deci\bel} gain.
Eccosorb filters are placed on both sides of the probe-fridge interface for both input and output side.
Additional Berkeley black coated copper shim enclosing the resonator packages is mounted to absorb stray photons to prevent quasiparticle generation.

The \textbf{NIST} fridge, Figure~\ref{fig:fridge}(b), is a FormFactor JDry-250 with a Radiall Cryogenic SP6T Ramses SMA \SI{18}{\giga\hertz} cryogenic switch that allows for the measurement of up to six samples at a time. As shown in the figure, the input line has \SI{60}{\deci\bel} worth of cryogenic attenuation as well as 5-10 \si{\deci\bel} from a strong IR filter (XMA 2482-5001CL) leading into the switch. After the sample, the amplification chain consists of, in order, a moderate IR filter (XMA 2482-5002CL), a travelling wave parametric amplifier (Lincoln Labs MIT TWPA), a high electron mobility transistor (Low Noise Factory HEMT), and a room-temperature low-noise amplifier (MITEQ AFS4 LNA). Samples are mounted on the \SI{10}{\milli\kelvin} stage cold finger inside a mu-metal shield, with circulators installed to allow characterization of all four $S$-parameters, and isolators to protect samples and amplifiers from upstream noise.

The \textbf{SU} fridge is an HPD Model 106 Shasta adiabatic demagnetization cryostat with a base temperature of \SI{100}{\milli\kelvin}.
Figure~\ref{fig:fridge}(c) describes the measurement setup using \SI{50}{\deci\bel} of room temperature attenuation and \SI{43}{\deci\bel} of cryogenic attenuation on the input and an Eccosorb filter that then leads into a HEMT and room temperature amplifier on the output.
We tested the SU cryostat with an isolator placed between the sample and the Eccosorb filter.
Measurements with and without the isolator gave similar results for the same \SI{60}{\second} BOE sample, as a result the isolator was removed for majority of the measurements to allow for loading more devices per cool down.

The \textbf{McMahon} fridge is a Bluefors LD dilution refrigerator with a base temperature of \SI{48}{\milli\kelvin}.
As described in Figure~\ref{fig:fridge}(d), the input lines have  \SI{60}{\deci\bel} attenuation from attenuators and \SI{7}{\decibel} attenuation from cabling inside the cryostat.
Copper Mountain M5180 VNA of 0 to \SI{-50}{\deci\bel m} dynamic range is connected to one programmable attenuator with tunability between 0 to \SI{60}{\deci\bel} attenuation at room temperature; the output lines host one high electron mobility transitor (HEMT) amplifier of \SI{37}{\deci\bel} gain, and one amplifier at room temperature of \SI{38}{\deci\bel} gain.
Eccosorb filters are placed on both sides of the probe-fridge interface for both input and output side.
Additional Berkeley black coated copper shim enclosing the resonator packages is mounted to absorb stray photons to prevent quasiparticle generation.

The \textbf{McMahon2} fridge is a Bluefors LD250 dilution refrigerator with a base temperature of \SI{19}{\milli\kelvin}.
As described in Figure~\ref{fig:fridge}(e), the input lines have \SI{60}{\deci\bel} attenuation from attenuators and around \SI{7}{\deci\bel} attenuation from the lines inside.
The input line has an Eccosorb filter and is mounted on a LNF circulator for isolation.
The output amplification line has an Eccosorb filter, along with a HEMT (Low Noise Factory, \SI{37}{\deci\bel} gain) and a room temperature amplifier (Mini circuits, \SI{38}{\deci\bel} gain).
Copper Mountain M5180 VNA of 0 to \SI{-50}{\deci\bel} dynamic range is connected to one programmable attenuator with tunability between 0 to \SI{60}{\deci\bel} attenuation at room temperature.

The \textbf{Fatemi} fridge is a Bluefors LD dilution refrigerator with a base temperature of \SI{10}{\milli\kelvin}.
As described in Figure~\ref{fig:fridge}(f), the input lines have \SI{60}{\deci\bel} attenuation from attenuators and around \SI{7}{\deci\bel} attenuation from the lines inside.
The output amplification line has an Eccosorb filter, along with a HEMT (Low Noise Factory, \SI{37}{\deci\bel} gain) and a room temperature amplifier (Mini circuits, \SI{38}{\deci\bel} gain).
Copper Mountain M5180 VNA of 0 to \SI{-50}{\deci\bel} dynamic range is connected to one programmable attenuator with tunability between 0 to \SI{60}{\deci\bel} attenuation at room temperature.
Eccosorb filters are placed on both sides of the probe-fridge interface for both input and output side.
Additional Berkeley black coated copper shim enclosing the resonator packages is mounted to absorb stray photons to prevent quasiparticle generation.

\begin{figure}
\centering
\includegraphics[width = 0.95\columnwidth]{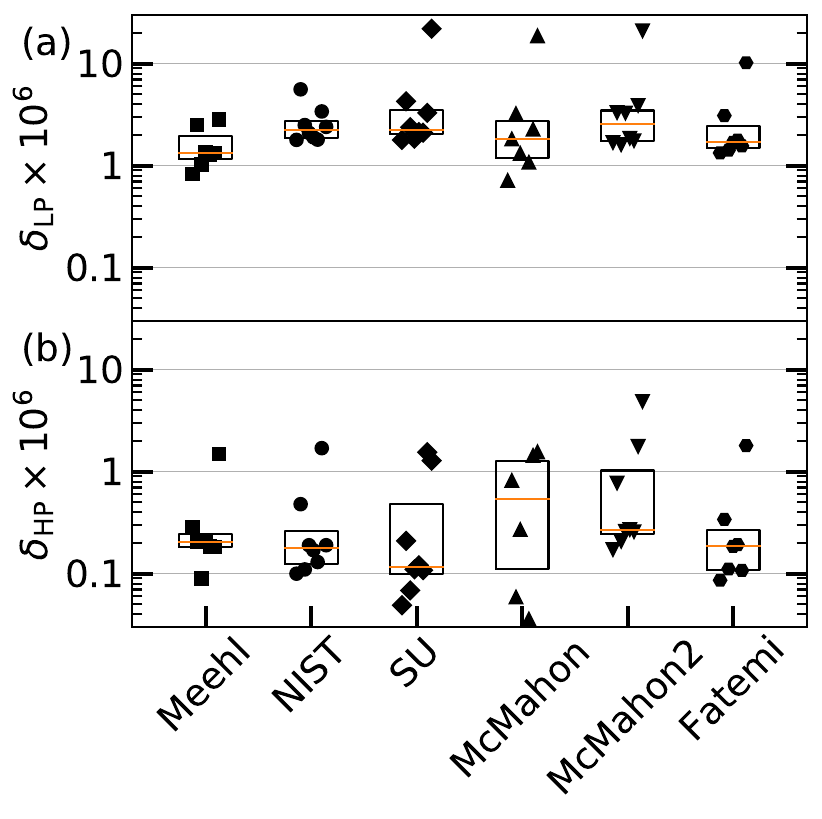}
\caption{Ta reference sample measured across the five fridges used in this experiment: Meehl, NIST, SU, McMahon, McMahon2, and Fatemi. Panel (a) shows $\delta_{LP}$, extracted low-power losses, and the panel (b) shows $\delta_{HP}$, high-power losses.
This is the same sample as in Jones, et al., sample LGS2~\cite{jones_grain_2023}.}
\label{fig:ta-calibration} 
\end{figure}

Prior to conducting measurements on our Nb films, the three setups were tested with a Ta resonator calibration sample with limited aging over time.
As shown in Figure~\ref{fig:ta-calibration}, the measurements across the three fridges showed good agreement in both LP and HP losses.
This is the same sample as in Jones, et al., sample LGS2~\cite{jones_grain_2023}.

\subsection{Resonator fits} \label{app:res_fits}
Measurements were conducted with five different setups as described in the previous section.
The complex transmission coefficient $S_{21}$ was measured with a vector network analyzer (VNA).
The $Q_{c}$ (coupled), $Q_{i}$ (internal), and $Q$ (loaded) quality factors were fitted using the diameter correction method (DCM)~\cite{mcrae_materials_2020,khalil_analysis_2012}:
\begin{equation}
    \label{eq:S21-fit}
    S_{21}(f) = 1-\frac{Q/\hat{Q_{c}}}{1+2iQ\frac{f-f_{0}}{f_{0}}},
\end{equation}
where $f$ is the frequency, $f_{0}$ is resonance frequency, and $\hat{Q_{c}}=Q_{c}e^{-i\phi}$ is the complex coupled quality factor.
A sample resonator fit is shown in Figure~\ref{fig:boxplots}(a).
The even distribution of measured points around the circle was achieved by using the "HPD" method as described in~\cite{baity_circle_2024}.
The average photon number was calculated using~\cite{mcrae_materials_2020}:
\begin{equation}
    \langle n\rangle=\frac{E_{0}}{\hbar\omega}=\frac{2}{\hbar\omega_{0}^{2}}\frac{Z_{0}}{Z_{r}}\frac{Q^{2}}{Q_{c}}P_\mathrm{app}.
\end{equation}
$P_\mathrm{app}$ is the power applied to the device, and $\omega_{0}$ is the resonant frequency.
$Z_{0}$ is the characteristic impedance of the microwave package the resonator chip is connected to, with a designed value of $\SI{50}{\ohm}$, $Z_{r}$ is the characteristic impedance of each superconducting resonator, also with a designed value of $\SI{50}{\ohm}$.
To extract the full performance of the resonators, we measured $S_{21}$ in the range of $10^{-1}$ to $10^{6}$ photons for $\langle n\rangle$.

We define loss $\delta_{i}$ as the inverse of the quality factor $Q_{i}$:
\begin{equation}
    \delta_{i}\approx\tan\delta_{i}=\frac{1}{Q_{i}}.
\end{equation}
HP losses are the average losses above $10^{5}$ photons, and LP losses are average losses below 1 photon.

\subsection{Temperature dependence}

\begin{figure}[b]
\centering
\includegraphics[width = 0.95\columnwidth]{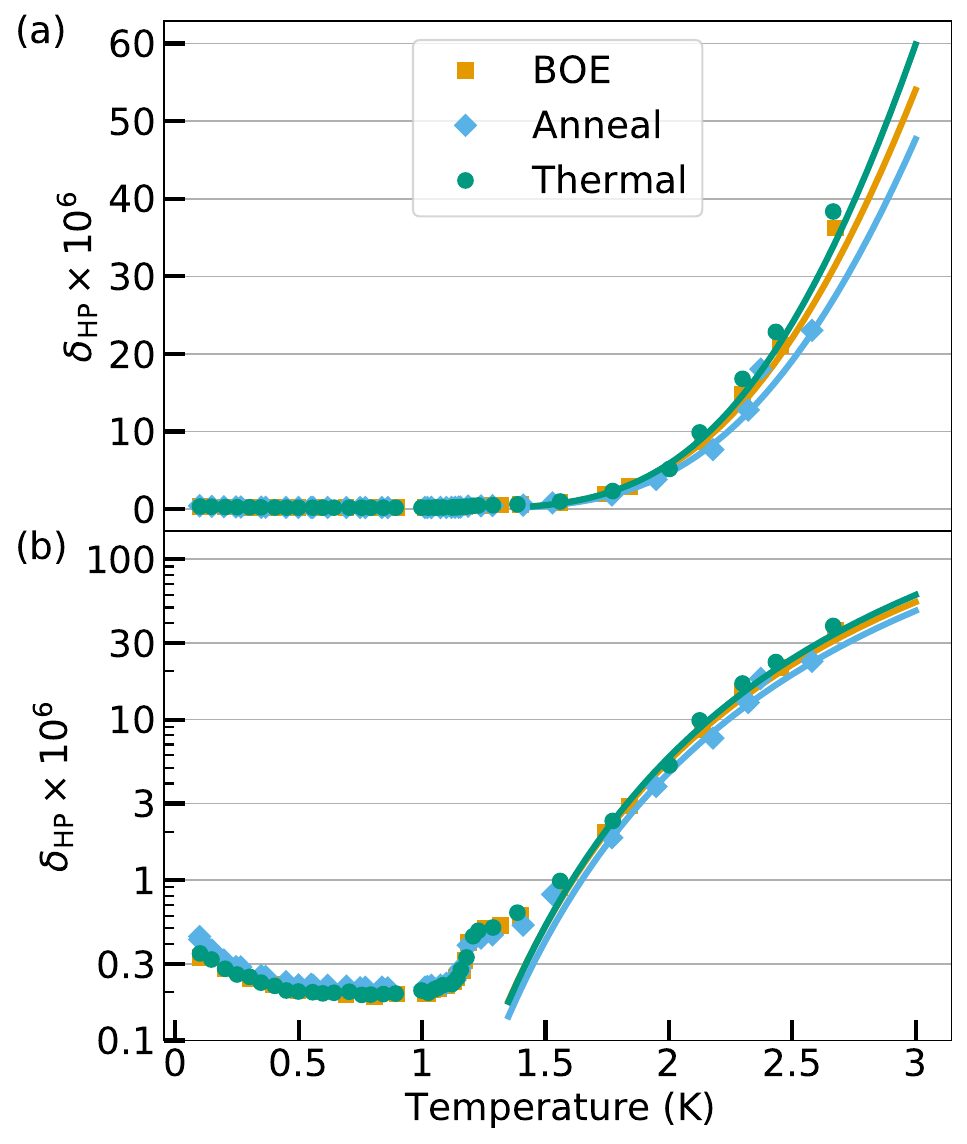}
\caption{ 
HP loss of three resonators prepared with DP1 and 1165, one from each sample preparation method (BOE, Anneal, and Thermal), plotted versus temperature (measured on the SU cryostat). 
The same data is plotted on a linear scale in panel (a) and on a log scale in panel (b).
Solid lines represent the Mattis-Bardeen (M-B) fit for quasi-particle loss at higher temperatures, as described in Eq.~\ref{eq:M-B}.
The values for $T_{c}$ from the M-B fits were: \SI{8.68}{\kelvin}, \SI{8.82}{\kelvin}, and \SI{8.85}{\kelvin}, for BOE, Anneal, and Thermal, respectively.
}
\label{fig:temperature} 
\end{figure}

We also measured the high-power internal quality factor of our Nb resonators as a function of temperature.
In Fig.~\ref{fig:temperature}, we show three sample resonators measured in the SU fridge.
The measurements were done at high photon numbers around 1$\times10^{5}$.
We fit the data beyond \SI{1.5}{\kelvin} with a Mattis-Bardeen (M-B) fit~\cite{mattis_theory_1958,chang_eliminating_2024,crowley_disentangling_2023} to extract an independent measurement of $T_{c}$ from the low-frequency electronic transport measurements:
\begin{equation}
    \delta_{i}(T)=\frac{\sinh(\frac{\hbar\omega}{2k_{B}T})K_{0}(\frac{\hbar\omega}{2k_{B}T})}{A_{QP}e^{\frac{\Delta}{k_{B}T}}},
    \label{eq:M-B}
\end{equation}
where, $\Delta=1.88k_{B}T_{c}$ is the effective superconducting gap for elemental Nb, $A_{QP}$ is the overall amplitude proportional to the kinetic inductance ratio, $k_{B}$ is the Boltzmann constant, and $K_{0}$ is the zero-th order modified Bessel function of the second kind.
The average $T_{c}$ for each set of samples is reported in Table~\ref{tab:substrates}.
We find that the $T_{c}$ of the three sample Nb resonators plotted in Figure~\ref{fig:temperature} are \SI{8.68}{\kelvin}, \SI{8.82}{\kelvin}, and \SI{8.85}{\kelvin}, for BOE, Anneal, and Thermal, respectively. 
The M-B $T_{c}$ values vary somewhat from the low-frequency measurements, which can be explained by the incomplete fit for the temperature dependent data. 

We now comment on the features below \SI{1.5}{\kelvin}.
We attribute the sharp increase in dissipation around \SI{1.1}{\kelvin} to the aluminum wire bonds used in packaging transitioning from superconducting to normal.
The large decrease in $\delta_{i}$ from low temperature to \SI{1}{\kelvin} is likely due to effects beyond the conventional quasiparticle and TLS models and remains under investigation~\cite{martinis_energy_2009}.

\balancecolsandclearpage

\bibliography{shared-references.bib}
\end{document}